%% file: main.tex
\documentclass{JFM-FLM_Au}

\usepackage{amsmath,bm,mathtools}
\usepackage{longtable,tabularx,multirow,array,threeparttable}
\IfFileExists{siunitx.sty}{%
  \usepackage{siunitx}%
}{%
  \providecommand{\SI}[2]{\ensuremath{##1\,##2}}%
  \providecommand{\micro}{\ensuremath{\mu}}%
  \providecommand{\metre}{\mathrm{m}}%
  \providecommand{\sisetup}[1]{}%
}
\usepackage{microtype}
\usepackage{subcaption}
\usepackage{xcolor}
\usepackage{enumitem}
\usepackage{placeins}
\usepackage{float}
\usepackage{pdflscape}
\usepackage[nameinlink,noabbrev]{cleveref}

\graphicspath{{figures/}}
\newcommand{\dd}{\mathrm{d}}
\newcommand{\Kn}{\mathit{Kn}}
\newcommand{\Ma}{\mathit{Ma}}
\newcommand{\Peff}{\mathcal{P}_{\mathrm{eff}}}
\newcommand{\eps}{\varepsilon}

\newcommand{\Cx}{C_x}
\newcommand{\Gzero}{\mathcal{G}_0}
\newcommand{\Gfull}{\mathcal{G}_{F}}
\newcommand{\Ep}{E_{p,\perp}}

\newcommand{\norm}[1]{\left\lVert #1\right\rVert}
\newcommand{\avg}[1]{\left\langle #1\right\rangle}
\newcommand{\devsym}[1]{\left[#1\right]^{(2)}}
\input{tables/headline_numbers_generated.tex}

\lefttitle{Ejtehadi \& Roohi}
\righttitle{Cross-stream pressure support and scalar relaxation}
\title{Cross-stream pressure support and the limits of scalar relaxation in rarefied Poiseuille flow}
\author{Omid Ejtehadi\aff{1} \and Ehsan Roohi\aff{2}}
\affiliation{\aff{1}School of Mechanical and Aerospace Engineering, Gyeongsang National University, 501 Jinjudaero, Jinju, Gyeongsangnam-do 52828, Republic of Korea
\aff{2}Department of Mechanical and Industrial Engineering, Riccio College of Engineering, University of Massachusetts Amherst, 160 Governors Drive, Amherst, MA 01003, USA}
\corresau{Ehsan Roohi, \href{mailto:roohie@umass.edu}{roohie@umass.edu}}

\hypersetup{hidelinks}
\AtBeginDocument{\hypersetup{
  pdftitle={Cross-stream pressure support and the limits of scalar relaxation in rarefied Poiseuille flow},
  pdfauthor={Omid Ejtehadi and Ehsan Roohi}}}

\begin{document}
\maketitle

\begin{abstract}
The weak wall-normal pressure variation in pressure-driven rarefied Poiseuille flow is a stringent test of higher-order constitutive models: it is almost invisible in the total-pressure norm, yet it is generated by anisotropic molecular stress. A tangent-form nonlinear coupled constitutive relation (NCCR) reproduces its convex topology, but the physical reason for its quantitative success remains unresolved. We ask whether the constitutive stress relation and reduced streamwise forcing are independently accurate or whether their effects compensate. A direct simulation Monte Carlo (DSMC) campaign is analysed using two-dimensional momentum budgets, a pressure-error norm based on the transverse signal, a matched-outlet-Knudsen comparison and componentwise tests of the pre-elimination NCCR balance. The non-equilibrium wall-normal stress over-supports the measured pressure defect, while streamwise transport of shear stress supplies an opposing correction. The state map, momentum budgets and forcing diagnostics show that neither outlet Knudsen nor outlet Mach number alone organises the pressure amplitude; along the fixed-ratio sequence, rarefaction is accompanied by larger changes in the constitutive diagnostics than in pressure amplitude. The DSMC-inferred stress relation departs from the fixed reduced coefficient, and even the best common scalar closes the dominant shear component far more accurately than the normal components that carry the pressure field. Correcting the coefficient alone can therefore worsen the reconstruction, whereas restoring omitted streamwise-momentum terms reduces the amplitude bias in strongly accelerated cases. The reduced law can remain accurate through stress--momentum compensation, showing that agreement of a weak non-equilibrium observable need not imply correct internal closure mechanics.
\end{abstract}

\section{Introduction}
\label{sec:introduction}
Pressure-driven gas transport through a microchannel is a canonical example of a flow whose macroscopic fields remain smooth while its molecular state departs from local thermodynamic equilibrium. When the channel height becomes comparable with the molecular mean free path, velocity slip and temperature jump are accompanied by Knudsen layers, normal-stress differences, tangential energy transport and altered mass-flow scaling \citep{ChapmanCowling1970,Kogan1969,Sone2000,SharipovSeleznev1998,EwartEtAl2007,GuEmerson2009,CorralCasasEtAl2022,AkhlaghiEtAl2023}. These effects are central to microfluidic and vacuum devices, but they also provide unusually clean tests of how continuum closures convert molecular transport into macroscopic fields.

Earlier DSMC studies of short micro/nanochannels showed that rarefaction alone does not characterise pressure-driven gas transport once compressibility and streamwise inertia become appreciable: under such conditions, low-Mach and long-channel reductions can misrepresent the pressure, velocity and mass-flow fields \citep{RoohiDarbandiMirjalili2009}. The present study revisits that rarefaction--compressibility interplay at the level of the weak cross-stream pressure field and its constitutive stress support.

The pressure variation across planar Poiseuille flow is a particularly demanding observable. A quasi-one-dimensional Navier--Stokes--Fourier (NSF) description makes the pressure uniform in the wall-normal direction, whereas molecular simulations and higher-order theories produce a weak convex profile with a centreline minimum \citep{Xu2003,Myong2011,RanaEtAl2016,JadhavEtAl2017,BrancherEtAl2021}. This transverse correction is small compared with the imposed streamwise pressure drop, but it is the direct macroscopic projection of stress anisotropy. It therefore exposes a limitation of conventional validation: a model may reproduce the total pressure extremely well while misrepresenting the small part of the field that contains the non-equilibrium physics.

Extended continuum descriptions recover non-zero normal stresses and tangential heat fluxes through different approximations, including Burnett-class models, regularised moment equations and kinetic closures \citep{OhwadaEtAl1989,StruchtrupTorrilhon2003,TaheriEtAl2009,TaheriStruchtrup2012,SinghEtAl2014,RathEtAl2021,YadavEtAl2024}. The nonlinear coupled constitutive relation (NCCR) is distinctive because its nonlinear stress coupling remains compact enough to admit analytical reductions \citep{Eu1992,Myong1999,Myong2014,Struchtrup2005}. For force-driven Poiseuille flow, \citet{Myong2011} used this structure to obtain a closed cross-stream pressure field. \citet{EjtehadiMyong2026} subsequently derived the pressure-driven tangent law and demonstrated that it reproduces the convex pressure topology over broad ranges of rarefaction, pressure ratio and channel aspect ratio; that work also established the associated entropy and non-Fourier transport behaviour.

That validation leaves a different question unresolved. The tangent law combines two reductions before the pressure profile is obtained: an algebraic relation between shear and normal stresses, and a quasi-one-dimensional approximation for the streamwise momentum forcing. Does the final pressure profile agree with molecular data because these two reductions are independently accurate, or because their effects compensate? This distinction is important not only for NCCR but for reduced higher-order models more generally, because transfer to another forcing or geometry depends on the internal balances, not solely on the accuracy of one final observable. Here we complement the earlier validation with DSMC statistics designed for complete two-dimensional momentum budgets, a signal-relative error norm and pre-elimination component tests.

The present study begins at this internal-mechanics level. A new DSMC campaign is organised into pressure-ratio and rarefaction sequences, a pair matched at outlet Knudsen number and a channel-length sequence at fixed height. The matched pair is used as a counterexample to a one-parameter rarefaction description; it is deliberately not matched in Mach number or acceleration history. The complete two-dimensional momentum equations are evaluated on the transverse-signal scale, so the normal-stress support can be separated from shear-stress transport and inertia. The reduced stress relation is then interrogated in two stages. First, a local coefficient is inferred as a diagnostic of the eliminated stress manifold; it is not introduced as a replacement closure. Second, the pre-elimination tensor equation is tested after fitting the best common scalar, which distinguishes a simple scalar-magnitude error from a componentwise failure of the normal stresses. Finally, a $2\times2$ set of DSMC-informed pressure reconstructions perturbs the constitutive coefficient and streamwise forcing separately and together.

The analysis reveals a coherent mechanism. Across the present sequences, local pressure forcing and compressible acceleration better organise the scale and downstream localisation of the transverse field than either outlet Knudsen or outlet Mach number alone, while rarefaction is more directly associated with changes in the stress response along the fixed-ratio branch. The wall-normal normal stress provides the leading mechanical support but systematically exceeds the measured defect; streamwise transport of shear stress offsets that excess. At the constitutive level, the dominant shear-weighted tensor direction remains aligned, yet one common scalar does not close the pressure-carrying normal components. The coefficient and forcing corrections consequently move the reconstructed pressure in opposite directions. The reduced tangent law can therefore remain accurate for compensating internal reasons. This provides a mechanistic explanation of its success while identifying where predictive extensions should focus.
\section{Problem formulation and reduced-law hypothesis}
\label{sec:framework}
We consider steady pressure-driven flow of a monatomic gas between two parallel isothermal plates. The observable of interest is the weak wall-normal pressure correction, not the dominant streamwise pressure drop. We denote the streamwise, wall-normal and spanwise Cartesian directions by $x$, $y$ and $z$, respectively; all mean fields are independent of $z$. The scalar pressure is $p$, the mass density is $\rho$, the macroscopic velocity vector is $\bm u$, and the non-equilibrium stress tensor is $\bm\Pi$; its Cartesian components are $\Pi_{ij}$, with $i,j\in\{x,y,z\}$. Centreline values carry the subscript $m$, so $p_m(x)$ and $\Pi_{ij,m}(x)$ are the centreline pressure and stress components, respectively. The centreline-referenced stress variation and dimensionless pressure defect are $\Delta\Pi_{ij}=\Pi_{ij}(x,y)-\Pi_{ij,m}(x)$ and $\delta_p=(p-p_m)/p_m$. All labels of the form P$k$ or N$k$ (for example, P1, N3 and P4) are DSMC case identifiers listed in \cref{tab:cases}; $k$ is the internal case number, and the prefixes P and N are campaign labels rather than physical variables. The centreline reference is not merely notational: $|\Pi_{yy,m}|/p$ exceeds the median transverse signal $|\Delta\Pi_{yy}|/p$ by factors of approximately 1.3, 4.4 and 8.4 in P1, N3 and P4, respectively, while the absolute $\Pi_{yy}$ changes sign across the channel. An absolute-stress formulation would therefore mix the weak transverse signal with a larger streamwise-varying offset and would be poorly conditioned for the present inference. The analysis follows the two reductions that connect the molecular stress field to the tangent pressure law: the reduced algebraic stress relation and the reduced streamwise forcing. Their effects are kept separate throughout. The coefficient introduced below is therefore a diagnostic probe of the constitutive layer, and the reconstructed pressure fields are conditional tests of the reduction rather than proposed autonomous continuum models.

\subsection{Geometry, observables and stress convention}
Let $L$ denote the channel length, $H$ the plate separation (channel height), and $\eps=L/H$ the channel aspect ratio. The planar channel occupies
\begin{equation}
0\le x\le L,\qquad -\frac{H}{2}\le y\le\frac{H}{2},\qquad \eps=\frac{L}{H}.
\label{eq:domain}
\end{equation}
All cases use $H=\SI{1.0}{\micro\metre}$; hence the three aspect ratios $\eps=4$, 8 and 16 correspond to $L=\SI{4}{\micro\metre}$, $\SI{8}{\micro\metre}$ and $\SI{16}{\micro\metre}$, respectively. These dimensions are those of the executed DSMC inputs.
The walls are isothermal and fully diffuse. Inlet and outlet pressures drive a steady monatomic gas. The molecular velocity is $\bm c=(c_x,c_y,c_z)$, the local macroscopic velocity is $\bm u=(u,v,0)$, and the peculiar molecular velocity is $\bm c'=\bm c-\bm u$, where $u$ and $v$ are the streamwise and wall-normal velocity components. Angular brackets $\avg{\cdot}$ denote a molecular ensemble average, $c'^2=\bm c'\boldsymbol\cdot\bm c'$ is the squared peculiar speed, and $\delta_{ij}$ is the Kronecker delta. The scalar pressure is one third of the trace of the molecular pressure tensor. We use the NCCR sign convention
\begin{equation}
\Pi_{ij}=\rho\avg{c_i'c_j'}-p\delta_{ij},
\qquad p=\frac13\rho\avg{c'^2},
\label{eq:stress-convention}
\end{equation}
Here $\rho$ is the mass density. With this convention the molecular momentum-flux tensor is $\rho\avg{\bm c'\bm c'}=p\bm I+\bm\Pi$, where $\bm I$ is the identity tensor. The symbol $\eta$ denotes the dynamic viscosity, $\nabla\bm u$ is the velocity-gradient tensor, and $\devsym{\cdot}$ denotes the symmetric trace-free (deviatoric) part of a second-order tensor; the NSF form $\bm\Pi=-2\eta\,\devsym{\nabla\bm u}$ then follows with the usual sign.

We write $\partial_x\equiv\partial/\partial x$ and $\partial_y\equiv\partial/\partial y$ for streamwise and wall-normal partial derivatives. The exact steady two-dimensional momentum equations are
\begin{align}
\rho\left(u\partial_xu+v\partial_yu\right)
+\partial_xp+\partial_x\Pi_{xx}+\partial_y\Pi_{xy}&=0,
\label{eq:x-momentum}\\
\rho\left(u\partial_xv+v\partial_yv\right)
+\partial_x\Pi_{xy}+\partial_y(p+\Pi_{yy})&=0.
\label{eq:y-momentum}
\end{align}
Integrating \cref{eq:y-momentum} from the centreline to $y$ gives
\begin{align}
p(x,y)-p_m(x)={}&-\Delta\Pi_{yy}(x,y)
-\int_0^y\rho\left(u\partial_xv+v\partial_yv\right)\dd y'
-\int_0^y\partial_x\Pi_{xy}\,\dd y',
\label{eq:integrated-y}
\end{align}
where $y'$ is a dummy wall-normal integration coordinate and $\dd y'$ denotes its differential; $\Delta\Pi_{ij}=\Pi_{ij}(x,y)-\Pi_{ij,m}(x)$ as defined above. This is the mechanically informative test. It separates normal-stress support from transverse inertia and streamwise transport of shear stress; it also avoids treating $p+\Delta\Pi_{yy}\simeq p_m$ as an independent comparison when $p$ and $\Pi_{yy}$ share molecular second moments.

\subsection{Reduced NCCR stress relation and tangent pressure law}
The pressure-driven reduction of \citet{EjtehadiMyong2026} uses
\begin{equation}
(\Delta\Pi_{xy})^2=-\frac32(p+\Delta\Pi_{yy})\Delta\Pi_{yy},
\label{eq:nccr-manifold}
\end{equation}
together with the approximation $\partial_y\Pi_{xy}\simeq-\partial_xp_m$. To interrogate this constitutive ingredient independently, we write the same algebraic structure as
\begin{equation}
(\Delta\Pi_{xy})^2=-C(p+\Delta\Pi_{yy})\Delta\Pi_{yy},
\label{eq:general-C}
\end{equation}
where the reduced NCCR value is $C=3/2$. The quantity $C$ is not proposed as a recalibrated universal closure coefficient. It is a diagnostic coordinate that measures how the molecular shear--normal-stress relation departs from the fixed reduced manifold. We use $x^*=x/L$ and $y^*=y/H$ for the dimensionless streamwise and wall-normal coordinates, and $\mathcal G(x)$ for a dimensionless local streamwise-forcing function. Given $p_m(x)$, the associated tangent law is
\begin{equation}
\frac{p(x,y)}{p_m(x)}=\sec^2\!\left[\frac{y/H}{\sqrt C}\,\mathcal G(x)\right],
\label{eq:tangent-law}
\end{equation}
with the reduced dimensionless forcing $\Gzero$
\begin{equation}
\Gzero=-H\frac{\partial\ln p_m}{\partial x}
=-\frac1\eps\frac{\partial\ln p_m}{\partial x^*}.
\label{eq:Gzero}
\end{equation}
For the measured arguments,
\begin{equation}
\delta_p\equiv\frac{p-p_m}{p_m}\simeq\frac{(y/H)^2\mathcal G^2}{C}.
\label{eq:small-defect}
\end{equation}
Because $H$ is fixed, changing $L/H$ by changing $L$ does not independently alter wall-normal confinement; it changes the realised streamwise forcing. The length sequence is therefore interpreted as an aspect-ratio-mediated forcing sequence.

The centreline form of \cref{eq:x-momentum} motivates
\begin{equation}
\Gfull=-\frac{H}{p_m}\left[
\frac{\dd(p_m+\Pi_{xx,m})}{\dd x}
+\rho_m u_m\frac{\dd u_m}{\dd x}
\right].
\label{eq:Gfull}
\end{equation}
Here $\rho_m$, $u_m$ and $\Pi_{xx,m}$ are the centreline density, streamwise velocity and streamwise normal-stress component, respectively, and $\Gfull$ is the centreline momentum-informed dimensionless forcing. This restores centreline inertia and the streamwise normal-stress gradient. It is a centreline momentum correction, not a complete two-dimensional closure.

\subsection{Diagnostic substitutions and error measures}
The two reduced ingredients can be perturbed independently. We denote the resulting pressure reconstructions by $\mathrm{M}_{0}$, $\mathrm{M}_{C}$, $\mathrm{M}_{G}$ and $\mathrm{M}_{CG}$, where the subscript identifies which ingredient is replaced relative to the original tangent law: $0$ denotes the unmodified reduction, $C$ the stationwise molecular stress coefficient $C_x(x)$, $G$ the centreline momentum-informed forcing $\mathcal G_F$, and $CG$ both substitutions together:
\begin{align}
\mathrm{M}_{0}:&\quad C=3/2,\quad \mathcal G=\Gzero,\nonumber\\
\mathrm{M}_{C}:&\quad C=\Cx(x),\quad \mathcal G=\Gzero,\nonumber\\
\mathrm{M}_{G}:&\quad C=3/2,\quad \mathcal G=\Gfull,\nonumber\\
\mathrm{M}_{CG}:&\quad C=\Cx(x),\quad \mathcal G=\Gfull.
\label{eq:diagnostic-models}
\end{align}
This $2\times2$ construction is the direct test of the compensation hypothesis. The comparison $\mathrm{M}_{C}$--$\mathrm{M}_{0}$ measures how the observable responds to the constitutive substitution, while $\mathrm{M}_{G}$--$\mathrm{M}_{0}$ measures its response to the momentum-forcing substitution. $\mathrm{M}_{CG}$ shows their joint action. All variants retain the DSMC centreline pressure; those containing $C$ also use a coefficient inferred from the same molecular clusters, and those containing $G$ use DSMC centreline inertia and $\partial_x\Pi_{xx,m}$. They are therefore conditional in-sample diagnostics, not predictive boundary-value models. Their purpose is not to rank four closures, but to establish whether the accuracy of $\mathrm{M}_{0}$ reflects two accurate reductions or opposing errors.

Let $p_{\mathrm{model}}$ denote a reconstructed pressure field and $p_{\mathrm{DSMC}}$ the pressure sampled by DSMC. The principal signal-normalised transverse-pressure error is $\Ep$, where the subscript $\perp$ denotes the wall-normal correction and $\norm{\cdot}_{\Omega}$ is the $L_2$ norm evaluated over the spatial region $\Omega$:
\begin{equation}
\Ep=\frac{\norm{p_{\mathrm{model}}-p_{\mathrm{DSMC}}}_{\Omega}}
{\norm{p_{\mathrm{DSMC}}-p_m}_{\Omega}},\qquad
\Omega=\{0.1\le x/L\le0.9,-1/2\le y/H\le1/2\}.
\label{eq:signal-error}
\end{equation}
The total-pressure error
\begin{equation}
E_{\mathrm{full}}=\frac{\norm{p_{\mathrm{model}}-p_{\mathrm{DSMC}}}}
{\norm{p_{\mathrm{DSMC}}}}
\label{eq:full-error}
\end{equation}
is retained for comparison with published NCCR validation. Here $E_{\mathrm{full}}$ is the relative total-pressure error and $\norm{\cdot}$ without a subscript denotes the same $L_2$ norm evaluated over the full computational domain. The two norms answer different questions. $E_{\mathrm{full}}$ measures the accuracy of the complete pressure field, which is dominated by $p_m(x)$; $\Ep$ measures the accuracy of the weak transverse correction that the higher-order stress relation is intended to represent.

\subsection{Pre-elimination algebraic NCCR compatibility}
Before eliminating the common scalar relaxation factor, the steady second-order NCCR stress equation has the algebraic form
\begin{equation}
\bm K\equiv2\devsym{\bm\Pi\boldsymbol\cdot\nabla\bm u}
+2p\devsym{\nabla\bm u}
=-\frac{p}{\eta}\bm\Pi\,q_{2nd}(\kappa),
\label{eq:nccr-K}
\end{equation}
where $\bm K$ is the retained symmetric trace-free kinematic tensor, the centred dot denotes a single tensor contraction, $q_{2nd}$ is the second-order NCCR scalar relaxation function, and $\kappa$ is its dimensionless non-equilibrium argument; specifically, $q_{2nd}(\kappa)=\sinh\kappa/\kappa$. In the present monatomic evaluation,
\begin{equation}
\kappa=\frac{(m k_B)^{1/4}T^{1/4}}{\sqrt{2}\,d\,p}
\left[\frac{\bm\Pi:\bm\Pi}{2\eta}
+\frac{\bm q\boldsymbol\cdot\bm q}{T k_{th}}\right]^{1/2},
\qquad k_{th}=\frac{\eta c_p}{\mathrm{Pr}},\quad c_p=\frac{5R}{2},\quad \mathrm{Pr}=\frac23,
\label{eq:kappa}
\end{equation}
where $m$, $d$ and $k_B$ are the molecular mass, VHS reference diameter and Boltzmann constant, $\bm q$ is the molecular heat flux, and $k_{th}$ is the thermal conductivity used in the diagnostic. The retained kinematic tensor $\bm K$ and the sampled stress tensor can then be compared without assuming the modelled scalar factor. The least-squares scalar multiplying $\bm\Pi$ is
\begin{equation}
\alpha_\star=\frac{\bm K:\bm\Pi}{\bm\Pi:\bm\Pi},\qquad
q_\star=-\frac{\eta}{p}\alpha_\star,
\label{eq:qstar}
\end{equation}
where $\alpha_\star$ is the least-squares best common scalar, $q_\star$ is the corresponding DSMC-required scalar relaxation factor, and the colon $:$ denotes the full $3\times3$ Frobenius double contraction, so symmetric off-diagonal entries contribute twice. The orthogonal fraction is
\begin{equation}
\chi_\perp=\frac{\norm{\bm K-\alpha_\star\bm\Pi}_F}{\norm{\bm K}_F}.
\label{eq:chi}
\end{equation}
Here $\norm{\cdot}_F$ is the Frobenius tensor norm and $\chi_\perp$ is the fraction of $\bm K$ orthogonal to the best scalar multiple of $\bm\Pi$. Small $\chi_\perp$ means that the tensor direction retained by the NCCR kinematics remains compatible with the sampled stress in the Frobenius norm. For reference, the prescribed-second-order component residual is
\begin{equation*}
R_{ij}^{(2)}=\frac{\operatorname{RMS}\!\left[K_{ij}+(p/\eta)\Pi_{ij}q_{2nd}\right]}{\operatorname{RMS}(K_{ij})},
\end{equation*}
where the superscript $(2)$ identifies use of the prescribed second-order relaxation factor. To test whether one common scalar also closes the pressure-carrying components, we evaluate
\begin{equation}
R_{ij}^{\star}=\frac{\operatorname{RMS}(K_{ij}-\alpha_\star\Pi_{ij})}{\operatorname{RMS}(K_{ij})},
\label{eq:Rstar}
\end{equation}
where $\operatorname{RMS}$ denotes the root-mean-square over the selected spatial points and $R_{ij}^{\star}$ is the dimensionless residual of component $ij$ after application of the best common scalar. For the pressure-relevant normal subspace we also define
\begin{equation}
R_N^\star=\frac{\operatorname{RMS}_F(\bm K_N-\alpha_\star\bm\Pi_N)}
{\operatorname{RMS}_F(\bm K_N)},
\qquad \bm K_N=\operatorname{diag}(K_{xx},K_{yy},K_{zz}),
\label{eq:RNstar}
\end{equation}
with $\bm\Pi_N$ defined analogously. The component and normal-subspace residuals are evaluated alongside the residual obtained with the prescribed $q_{2nd}$. The ratio $q_\star/q_{2nd}$ measures the scalar-magnitude mismatch, whereas $R_{ij}^{\star}$ and $R_N^\star$ isolate the incompatibility that remains after the best common scalar is applied. This is a compatibility test of the closed algebraic NCCR balance, not a direct measurement of every term in the unclosed molecular stress-moment equation.

\section{DSMC campaign and statistical inference}
\label{sec:methods}
\subsection{Gas model, boundaries and sampled moments}
The working gas is monatomic argon at $T_{in}=T_w=\SI{300}{K}$, where $T_{in}$ and $T_w$ are the inlet-gas and wall temperatures. The variable-hard-sphere (VHS) model uses molecular mass $6.63\times10^{-26}\,\mathrm{kg}$, reference diameter $4.17\times10^{-10}\,\mathrm{m}$ specified at $T_{ref}=\SI{273}{K}$, and viscosity--temperature exponent $\omega=0.81$. Constitutive post-processing uses $\eta=2.23\times10^{-5}(T/\SI{300}{K})^{0.81}\,\mathrm{Pa\,s}$. All simulations were performed with the in-house \texttt{DSMC2} solver. The code descends from the solver used for rarefied shear-driven micro/nanoflows and second-law Couette calculations \citep{EjtehadiRoohiEsfahani2012,EsfahaniEjtehadiRoohi2013}; its present pressure-boundary implementation and channel configuration were verified against slip-flow analytical solutions and high-Knudsen DSMC reference data in Appendix~A of \citet{EjtehadiMyong2026}. Molecular collisions use the no-time-counter algorithm in the standard operator-splitting DSMC framework \citep{Bird1994,Wagner1992}. The walls are diffuse with complete thermal and momentum accommodation. Pressure boundaries follow the characteristic-based inlet procedure of \citet{WangLi2004} and the subsonic outlet formulation of \citet{LiouFang2000}. Related DSMC pressure-boundary and reservoir formulations are discussed by \citet{IkegawaKobayashi1990,FarbarBoyd2014,ShahEtAl2018}, while \citet{AokiEtAl2022} provide a modern kinetic derivation of gas--surface boundary conditions for the Boltzmann equation. The sampled fields include mass density $\rho$, streamwise and wall-normal velocities $u$ and $v$, temperature $T$, scalar pressure $p$, all independent components of $\bm\Pi$, the streamwise and wall-normal heat-flux components $q_x$ and $q_y$, and selected third- and fourth-order moments where available. Because the grid has an even number of wall-normal cells, every centreline value is the average of the two innermost cell-centre values after parity enforcement.

The imposed pressure values are those in the executed \texttt{case.nml} files. We write $P_{\mathrm{in}}$ and $P_{\mathrm{out}}$ for the prescribed inlet and outlet reservoir pressures, so $P_{\mathrm{in}}/P_{\mathrm{out}}$ is the nominal pressure ratio. Let $P_{\mathrm{in}}^{monitor}$ and $P_{\mathrm{out}}^{monitor}$ denote the corresponding solver-monitored inlet and outlet pressures, and let $\avg{\cdot}_{\mathrm{final}\ 20\%}$ denote a time average over the final 20\% of the recorded convergence history. The realised effective pressure ratio $\Peff$ is
\begin{equation}
\Peff=\frac{\avg{P_{\mathrm{in}}^{monitor}}_{\mathrm{final}\ 20\%}}
{\avg{P_{\mathrm{out}}^{monitor}}_{\mathrm{final}\ 20\%}}.
\label{eq:Peff}
\end{equation}
The local Knudsen number is $\Kn=\lambda/H$, where $\lambda$ is the solver-reported local VHS mean free path. The outlet value $\Kn_{\mathrm{out}}$ is the cross-sectional mean of $\Kn$ at the last sampling station; the outlet centreline value is retained separately. The reported outlet-centreline Mach number is $\Ma_{\mathrm{out},c}=u_c/\sqrt{\gamma R T_c}$, where the subscript $c$ denotes the centreline at the last sampling station, $u_c$ and $T_c$ are the corresponding streamwise velocity and temperature, $R$ is the specific gas constant of argon, and $\gamma=5/3$ is the ratio of specific heats for a monatomic gas. The centreline subscript $c$ in this Mach-number definition is unrelated to the molecular velocity $\bm c$. More generally, the local Mach number is $\Ma(x,y)=\lvert\bm u(x,y)\rvert/\sqrt{\gamma R T(x,y)}$, where $\lvert\bm u\rvert$ is the speed magnitude.

\subsection{Operating points and provenance}
\Cref{tab:cases} summarises the nine physical cases. Each P case comprises three independent cold-start seeds. Each N case comprises five consecutive, non-overlapping sampling blocks formed by restarting the same calibrated trajectory with fresh sampling accumulators; these blocks are finite-cluster units, not independent cold starts. Thus P and N are internal case-prefix labels for the two statistical constructions, not abbreviations for physical regimes. All molecular fields were generated in the present campaigns; no raw field, figure or table from \citet{EjtehadiMyong2026} is reused. Supplementary table~S10 documents the sampling grid, collision subcells, initial particles per cell, time step, number and type of statistical clusters, averaging duration per cluster and total accumulated sampling.

\begin{table}
\centering
\caption{Operating points and primary molecular diagnostics. All cases use $H=\SI{1.0}{\micro\metre}$ and $L=\eps H$. The defect is the two-wall average at the common cell-centre location $|y/H|=0.491667$ over $0.1\le x/L\le0.9$. Coefficient intervals are 95\% finite-cluster percentile intervals. P cases use three independent seeds and N cases five consecutive non-overlapping blocks; full computational settings are given in supplementary table~S10. Case labels retain the internal campaign identifiers for traceability, so the numbering is not contiguous. The numerical replicas V1 and V2 are excluded from the nine physical cases. N5 is retained as a low-amplitude long-channel endpoint, but no $C_x$ value is reported because its interior-mean wall defect is only $0.010\%$ and it is not used in the constitutive conclusions.}
\label{tab:cases}
\resizebox{\textwidth}{!}{\input{tables/operating_points_generated.tex}}
\end{table}

The campaign supplies three complementary comparisons: (i) P1--P2--N1 increase realised pressure ratio at nearly fixed inlet pressure and $\eps=8$; (ii) P1--N2--N3--P4 increase rarefaction at nominal pressure ratio $2.5$ and $\eps=8$; (iii) P2/N2 match $\Kn_{\mathrm{out}}$ while changing realised pressure ratio, outlet Mach number and acceleration history. The N4--P1--N5 sequence changes channel length at fixed height and imposed pressure ratio, and is interpreted through its realised local forcing.

\subsection{Resolution and convergence}
The baseline sampling grid is $100\times60$ for $\eps=8$, $50\times60$ for $\eps=4$ and $200\times60$ for $\eps=16$. Collision selection uses $2\times2$ subcells inside every sampling cell. We denote the streamwise and wall-normal collision-subcell dimensions by $\Delta x_{sub}$ and $\Delta y_{sub}$, the DSMC time step by $\Delta t$, and the local mean collision time by $\tau$. Across all physical cases,
\begin{equation}
\max\frac{\Delta x_{sub}}{\lambda}=\MaxDxSubLambda{},\qquad
\max\frac{\Delta y_{sub}}{\lambda}=\MaxDySubLambda{},\qquad
\max\frac{\Delta t}{\tau}=\MaxDtTau{}.
\label{eq:resolution-ratios}
\end{equation}
The distinction between sampling cells and collision subcells is essential in the dense inlet region. Cell-size, time-step and particle-sampling effects follow the verification principles established for stochastic particle algorithms \citep{AlexanderEtAl1998,Hadjiconstantinou2000,HadjiconstantinouEtAl2003,MyongEtAl2019}. Collision subcells and adaptive collision strategies provide an additional route to controlling dense-region discretisation without conflating sampling cells with collision cells \citep{StefanovEtAl2022}. Two numerical verification variants of P2 are also used: V1 doubles the sampling-grid resolution in both directions and V2 halves the time step. They are numerical replicas of P2 and are not counted among the nine physical cases in \cref{tab:cases}. Near-wall observables are compared at the common dimensionless wall-normal position $|y^*|=0.491667$, where $y^*=y/H$, not at mesh-dependent first-cell centres.

Interior mass-flux variation computed directly from the sampled fields is below $2.1\times10^{-3}\%$ in the production cases. As an independent check, the ratio of the solver boundary mass-flux counter to a direct $\int \rho u\,\dd y$ integral of the sampled interior fields has a mean of 1.0053 across the production and verification cases for which both counters are available, with 0.20\% case-to-case scatter. The larger inlet/outlet counter mismatch is therefore treated as a pressure-boundary accounting diagnostic rather than an interior conservation error. The high-rarefaction case P4 is additionally analysed using boundary distances measured in local centreline mean free paths.

\subsection{Derivatives, parity and exact budgets}
Two-wall symmetry is imposed before differentiation: scalar variables and normal stresses are even; $v$, $\Pi_{xy}$ and $q_y$ are odd. Streamwise derivatives use a third-order Savitzky--Golay polynomial \citep{SavitzkyGolay1964} over a window proportional to streamwise grid size; half- and double-window tests are retained. Wall-normal derivatives use second-order non-uniform finite differences on the cell-centre locations. The integrated wall-normal budget is recomputed from each statistical cluster resample, including smoothing, differentiation and centreline integration.

The default statistical region is $0.1\le x/L\le0.9$. For the boundary-aware masks, $d_{in}$ and $d_{out}$ denote the streamwise distances from a sampling station to the inlet and outlet pressure boundaries, $\lambda_m$ is the local centreline mean free path, and $n$ is the prescribed minimum number of mean free paths. Boundary-aware variants intersect the default window with
\begin{equation}
\frac{d_{in}}{\lambda_m}>n,\qquad \frac{d_{out}}{\lambda_m}>n,
\qquad n=1,2,3.
\label{eq:mfp-mask}
\end{equation}
Thus every stricter boundary mask is nested inside the default interior.

\subsection{Finite-cluster bootstrap and order-two stress products}
The statistical unit is an entire cold-start seed or an entire non-overlapping averaging block; spatial points are never resampled independently. Cluster resampling follows the non-parametric bootstrap framework of \citet{EfronTibshirani1993}, while the nonlinear unbiased products below are order-two U-statistics in the sense of \citet{Hoeffding1948}. This field-level uncertainty treatment complements recent kinetic studies that quantify model and relaxation-rate uncertainty in rarefied molecular gases \citep{LiEtAl2021UQ}. In this subsection, $n$ denotes the number of statistical clusters and is distinct from the boundary-mask integer $n$ in \cref{eq:mfp-mask}. For $n$ clusters, the ordinary cluster bootstrap draws $n$ clusters with replacement. Because $n=3$ or $5$, all unique multinomial count vectors are enumerated exactly: 10 vectors for $n=3$ and 126 for $n=5$, with their exact multinomial probabilities. Exact enumeration removes Monte Carlo variability in evaluating this finite-cluster resampling distribution; it does not increase the number of physical seeds or sampling blocks represented by the data.

Centreline subtraction, parity enforcement, selection, differentiation, integration and pressure reconstruction are repeated inside every resample. The nonlinear products defining the stress relation use order-two U-statistics \citep{Hoeffding1948}. Here $r,s\in\{1,\ldots,n\}$ index the statistical clusters; $A_r$, $U_r$ and $V_r$ are generic cluster-level estimates; $\mu_A$, $\mu_U$ and $\mu_V$ are their corresponding ensemble means; a hat denotes an estimator; and the sums over $r\ne s$ include all ordered pairs of distinct clusters:
\begin{align}
\widehat{\mu_A^2}&=\frac{\sum_{r\ne s}A_rA_s}{n(n-1)},\\
\widehat{\mu_U\mu_V}&=\frac{\sum_{r\ne s}U_rV_s}{n(n-1)}.
\label{eq:ustats}
\end{align}
These expressions remove the leading positive square bias and product covariance bias while preserving the common-centreline covariance. The dimensionless stress-product variables $X$ and $Y$ are defined by
\begin{equation}
X=-\frac{(p+\Delta\Pi_{yy})\Delta\Pi_{yy}}{p_m^2},\qquad
Y=\frac{(\Delta\Pi_{xy})^2}{p_m^2},
\label{eq:XY}
\end{equation}
The stationwise coefficient $\Cx$ is the total-least-squares (TLS) slope through the origin at each streamwise station, using upper-half points satisfying
\begin{equation}
|y/H|\ge0.1,\qquad X>3\,\mathrm{SE}_{JK}(X),\qquad Y\ge0.
\label{eq:selection}
\end{equation}
Here $\mathrm{SE}_{JK}(X)$ is the jackknife standard error of $X$ across the statistical clusters. The case summary is the median over stations. Global TLS, global ordinary-least-squares (OLS), reverse OLS and median $Y/X$ are reported as alternative estimands. Cluster-bootstrap intervals quantify sampling uncertainty in the complete analysis chain; they do not quantify model-form error or a boundary-condition bias shared by all clusters. For the P-series cases, the coefficient conclusion is effect-size driven rather than dependent on fine interval resolution: despite using three independent seeds, the complete finite-cluster interval of every resolved P case remains below $C=3/2$, with the largest upper endpoint equal to 1.360.

\subsection{Reference baselines}
The reduced flat-pressure NSF baseline has $p=p_m$ and hence $\Ep=1$ by construction. This baseline is intentionally narrower than solved slip-flow and second-order-slip models such as those of \citet{ArkilicEtAl1997,BeskokKarniadakis1999,MaurerEtAl2003,DongariEtAl2009}; it isolates the transverse pressure mechanism rather than overall microchannel conductance. A less restrictive a priori first-order comparison evaluates
\begin{equation}
\Pi_{yy}^{NSF}=-2\eta\left(\partial_yv-\frac13\nabla\boldsymbol\cdot\bm u\right)
\label{eq:nsf-piyy}
\end{equation}
where $\Pi_{yy}^{NSF}$ is the NSF estimate of the wall-normal non-equilibrium stress and $\nabla\boldsymbol\cdot\bm u=\partial_xu+\partial_yv$ is the velocity divergence. The comparison reconstructs $p_m-(\Pi_{yy}^{NSF}-\Pi_{yy,m}^{NSF})$, with $\Pi_{yy,m}^{NSF}$ denoting its centreline value. This is a local constitutive test, not an independently solved two-dimensional NSF/slip boundary-value problem.

For context, two present operating points are near those used in Appendix B of \citet{EjtehadiMyong2026}. Because their raw molecular fields are unavailable, the comparison uses published full-domain errors and present full-domain metrics; it is explicitly near matched rather than a common-window reprocessing.

\section{Results}
\label{sec:results}
The results proceed from the measured mechanics to the internal model diagnosis. We first establish the complete wall-normal support of the pressure field and the streamwise terms omitted by the reduced forcing. Controlled case sequences then identify the coordinate that organises pressure amplitude. The stress relation is subsequently examined at the eliminated-manifold and pre-elimination component levels. The final pressure reconstructions combine these findings to determine whether the two reduced ingredients reinforce or compensate one another.

\subsection{Physical anatomy of the pressure-support mechanism}
\begin{landscape}
\begin{figure}[p]
\centering
\includegraphics[width=0.97\linewidth]{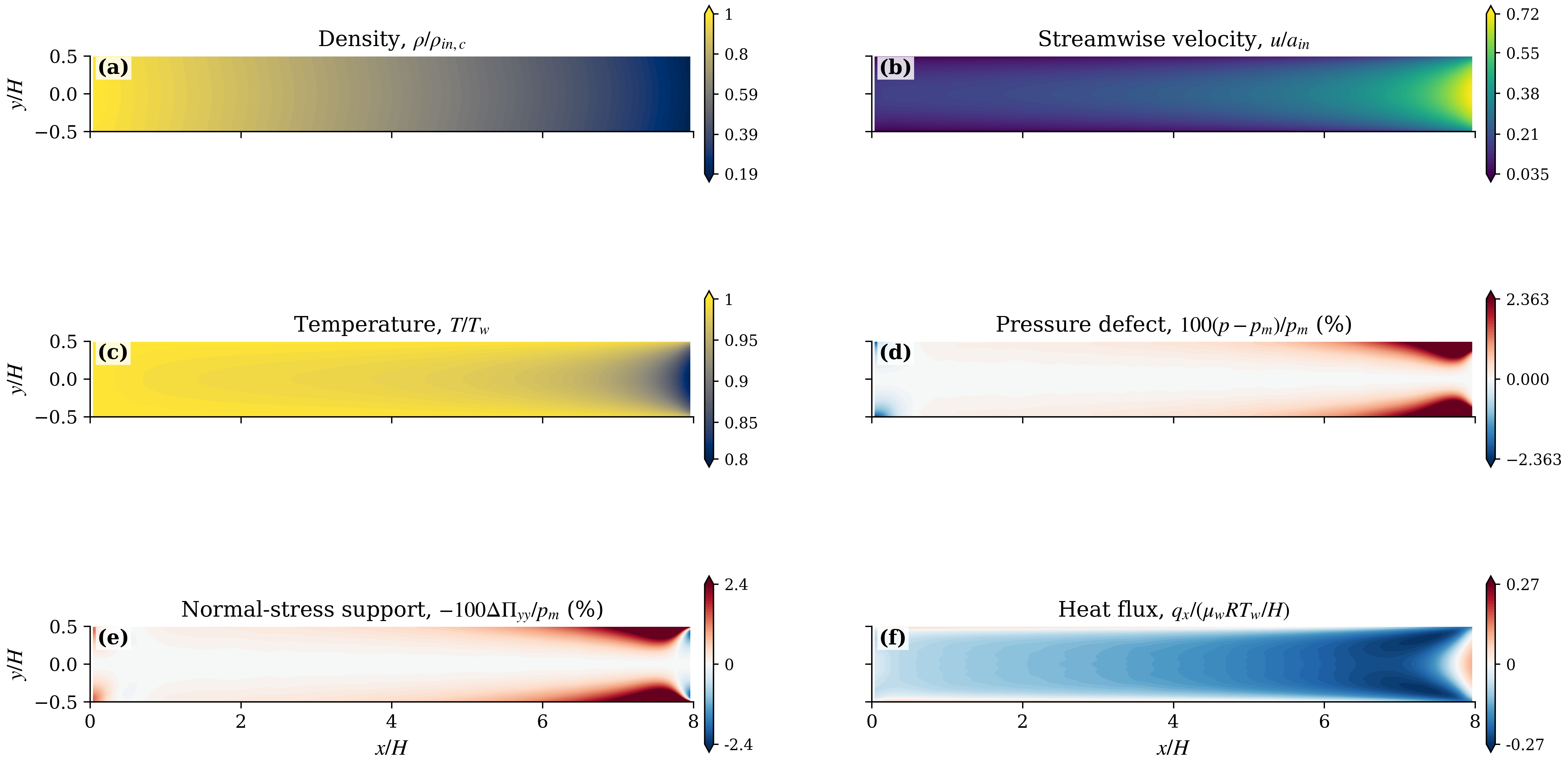}
\caption{Physical anatomy of case N1 in the actual channel coordinates $0\le x/H\le8$ and $-0.5\le y/H\le0.5$.  Panels show (a) density, (b) streamwise velocity, (c) temperature, (d) transverse pressure defect, (e) wall-normal normal-stress support and (f) streamwise heat flux.  Here $\rho_{in,c}$ is the inlet-centreline density, $a_{in}=\sqrt{\gamma R T_{in}}$ is the inlet speed of sound, and $\mu_wRT_w/H$ is the heat-flux scale, with $\mu_w$ denoting the dynamic viscosity evaluated at the wall temperature.  Panels (d) and (e) use the same diverging colour limits, making the close spatial correspondence between the pressure defect and $-\Delta\Pi_{yy}/p_m$ directly visible.  The downstream localisation of both fields coincides with the strongest acceleration.  A light sub-cell Gaussian interpolation is used only for contour rendering; all statistics, derivatives and profiles use the unsmoothed DSMC fields.}
\label{fig:anatomy}
\end{figure}
\end{landscape}

\Cref{fig:anatomy} shows N1, a strong but subsonic pressure-driven case.  Density decreases and the gas accelerates towards the outlet, while the temperature falls because the pressure work and non-equilibrium energy transport are not locally balanced by an isothermal response.  The transverse pressure defect is much smaller than the imposed longitudinal pressure change, but it forms a coherent even-in-$y$ field with a centreline minimum and wall maxima.  Its downstream growth identifies the region in which the molecular stress anisotropy is most strongly expressed.

The map of $-\Delta\Pi_{yy}/p_m$ occupies almost the same region and has the same wall-normal symmetry, immediately suggesting a normal-stress support mechanism.  This similarity is physically meaningful but does not by itself establish a complete balance, because pressure and normal stress share molecular second moments and because streamwise transport of shear stress also enters the exact wall-normal momentum equation.  The upstream-directed $q_x$ is included only to complete the flow anatomy; the non-Fourier heat-transport mechanism has been analysed comprehensively elsewhere \citep{AkhlaghiEtAl2018,JohnEtAl2013,EjtehadiMyong2026} and is not claimed here as a new result.

The small magnitude of the signal motivates the error norm $\Ep$.  A total-pressure error is dominated by the centreline variation $p_m(x)$, whereas $\Ep$ asks whether a model reproduces the much smaller transverse component represented by panels (d) and (e) of \cref{fig:anatomy}.  This distinction becomes essential when a total-field error of order $10^{-3}$ coexists with a transverse-signal error of order $10^{-1}$.

\subsection{How the transverse pressure is supported}
\begin{landscape}
\begin{figure}[p]
\centering
\includegraphics[width=0.97\linewidth]{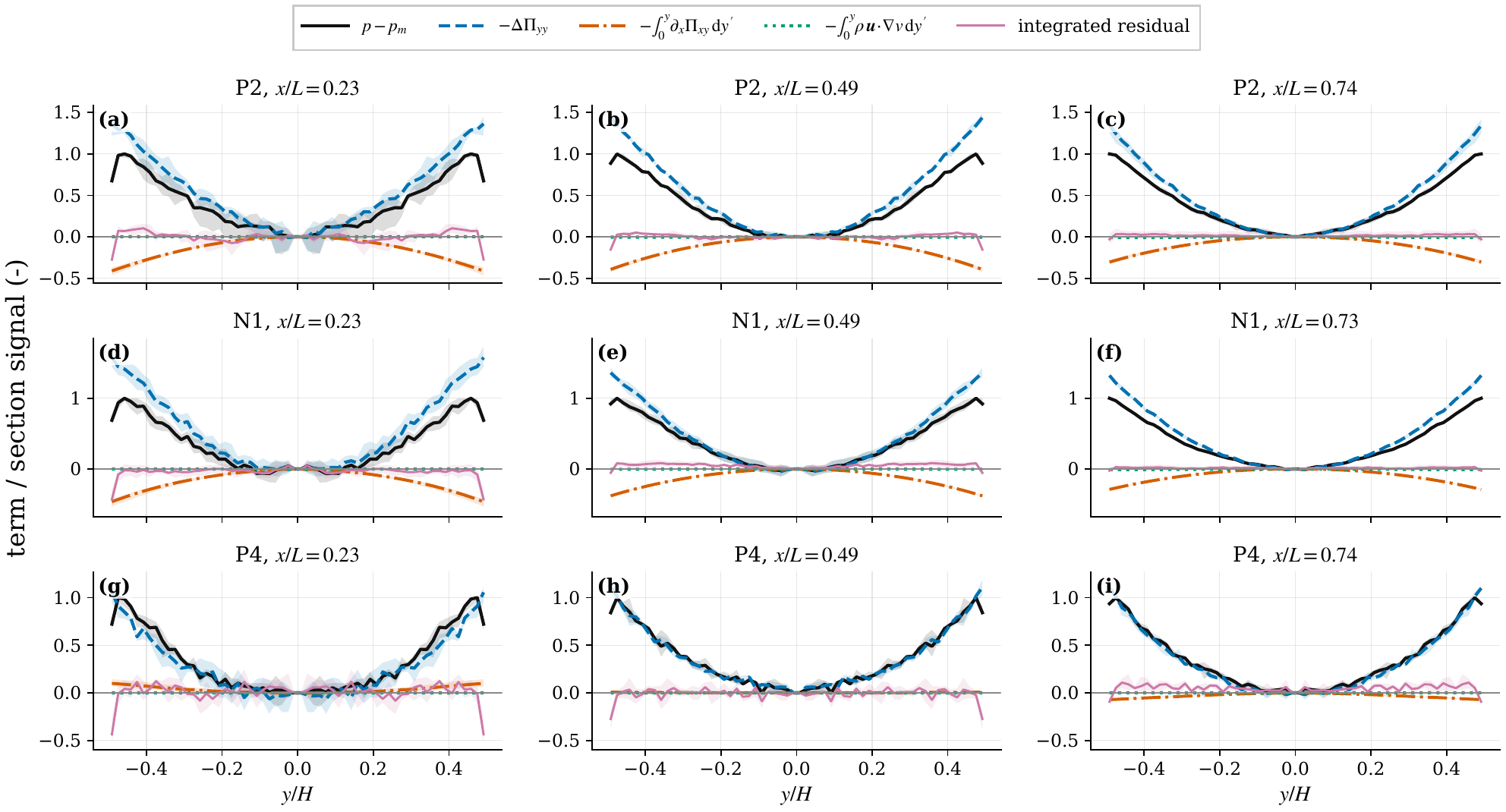}
\caption{Evaluation of the integrated wall-normal momentum budget for P2, N1 and P4 at $x/L\simeq0.245$, 0.495 and 0.745.  The black curve is the measured pressure defect.  The normal-stress contribution supplies the dominant positive support, the integrated streamwise shear-stress transport supplies the principal opposing correction and transverse inertia remains small.  Every term is dimensionless and normalised by the maximum pressure signal at the same cross-section.  Shaded bands denote 95\% finite-cluster percentile intervals.}
\label{fig:ybudget}
\end{figure}
\end{landscape}

\Cref{fig:ybudget} evaluates the integrated balance \cref{eq:integrated-y} for P2, N1 and P4 at three streamwise stations.  Each term is divided by the maximum measured pressure signal at the corresponding section, so the ordinate measures the importance of a term relative to the phenomenon being explained rather than relative to the much larger centreline pressure.  The normal-stress contribution has the same convex shape as the pressure signal but generally exceeds its amplitude.  Integrated streamwise transport of shear stress has the opposite sign and removes the excess support.  Transverse inertia is small throughout the selected region.

The balance develops systematically along the channel.  Upstream, all transverse terms are weak because the gas has undergone little acceleration.  Downstream, the pressure defect and $-\Delta\Pi_{yy}$ grow together, while the shear-transport correction becomes large enough to be visible on the signal scale.  P4 retains the same hierarchy of terms, although its profiles are statistically rougher and more sensitive to the distance from the pressure boundary.

Writing the measured signal as $s=p-p_m$ and an integrated contribution as $b_k$, the reported signed projection is $B_k=\langle b_k,s\rangle_\Omega/\langle s,s\rangle_\Omega$ and the residual is $\|s-\sum_k b_k\|_\Omega/\|s\|_\Omega$, using equal cell-centre weights after two-wall symmetrisation. Case-level values are reported in \cref{tab:budget}. In resolved cases the normal-stress projection is \NormalSupportMin{}--\NormalSupportMax{}, while the shear-transport projection is negative with magnitude \ShearOppMin{}--\ShearOppMax{}. The integrated residual is \YResidualMin{}--\YResidualMax{} of the transverse signal and is retained explicitly rather than interpreted as precision closure. Spatial residual maps and boundary-mask sensitivity are reported in supplementary figures~S2--S3 and \cref{fig:sensitivity}; finite-cluster intervals quantify sampling variation but not a smoothing or pressure-boundary bias common to every cluster.

\begin{table}
\centering
\caption{Finite-cluster evaluation of the integrated wall-normal momentum budget.  Normal-stress, shear-transport and transverse-inertia entries are signed inner-product projections on the measured pressure signal; the residual is a ratio of Euclidean $L_2$ norms.  Brackets give 95\% finite-cluster percentile intervals.  Values greater than unity for the normal-stress projection mean that $-\Delta\Pi_{yy}$ over-supports the measured defect and must be offset by the negative shear-transport contribution.}
\label{tab:budget}
\resizebox{\textwidth}{!}{\input{tables/budget_ci_generated.tex}}
\end{table}

The result changes the mechanical interpretation of the reduced identity $p-p_m\simeq-\Delta\Pi_{yy}$. The normal stress does not merely correlate with the pressure field; it systematically over-supports it. The missing streamwise transport of shear stress has the same even wall-normal symmetry and removes the excess without strongly altering the convex shape. This momentum-balance compensation explains how the tangent topology is preserved while its amplitude retains a substantial signal-level error.

\subsection{When the reduced streamwise forcing breaks down}
The second ingredient of the tangent law is the replacement of the complete streamwise momentum balance by $\partial_y\Pi_{xy}\simeq-\partial_xp$. Supplementary table~S1 quantifies what is removed by this step. All RMS terms are normalised by RMS $|\partial_xp|$, so their magnitudes can be compared on the pressure-gradient scale. The pressure gradient and shear-stress gradient form the leading balance in every case, but the neglected terms are not uniformly small. Convective inertia remains secondary along the low-Mach rarefaction ladder and rises to approximately 10--15\% in P2, N1 and N4. The streamwise normal-stress gradient is typically one to two percent of the pressure-gradient scale; although smaller, it is comparable with the weak transverse correction that the reduced law seeks to predict.

The outlet Mach number is a compact indicator of the resulting compressible state, but it is not the term that enters the momentum balance directly. A more mechanical local measure is the centreline inertial fraction
\begin{equation}
\mathcal I_m(x)=\frac{\left|\rho_m u_m\,\dd u_m/\dd x\right|}{\left|\dd p_m/\dd x\right|},
\label{eq:inertial-fraction}
\end{equation}
which compares centreline convective acceleration with the pressure-gradient forcing. The correction from $\Gzero$ to $\Gfull$ incorporates this inertial contribution, together with $\partial_x\Pi_{xx,m}$; it is therefore more mechanistic than Mach number alone.

The consequence is a separation between rarefaction and streamwise inhomogeneity. Along P1--N2--N3--P4, $\Ma_{\mathrm{out},c}$ decreases while the station-typical stress coefficient increases, so the constitutive departure on that branch cannot be explained by growing convective inertia. A direct control on the shear equation leads to the same conclusion. Defining the dimensionless dominant-shear kinematic factor $\kappa_s=K_{xy}/[\dot\gamma(p+\Pi_{yy})]$, where $K_{xy}$ is the shear component of $\bm K$ and $\dot\gamma=\partial_yu$ is the local shear rate, its canonical-interior value ranges from 0.997 to 0.999 across P1, N3 and P4. Thus the shear kinematics remain within 0.3\% of the one-dimensional relation even as the normal-component incompatibility and required scalar relaxation change. Conversely, N4 carries a strong inertial contribution at comparatively low $\Kn$. The momentum-informed forcing $\Gfull$ is introduced to diagnose this second reduction. It restores the centreline terms omitted by $\Gzero$ but does not replace the complete two-dimensional balance established in \cref{eq:x-momentum}.

\subsection{Matched outlet rarefaction but different dynamical states}

The pair P2--N2 tests whether the transverse-pressure field can be parameterised by outlet rarefaction alone. The two cases have almost identical outlet Knudsen numbers,
\[
\Kn_{\mathrm{out}}^{P2}=0.218,
\qquad
\Kn_{\mathrm{out}}^{N2}=0.219,
\]
but they are not dynamically matched. Their realised pressure ratios are
\[
\Peff^{P2}=4.864,
\qquad
\Peff^{N2}=2.517,
\]
and their outlet centreline Mach numbers are
\[
\Ma_{\mathrm{out},c}^{P2}=0.649,
\qquad
\Ma_{\mathrm{out},c}^{N2}=0.236.
\]
Thus, the outlet Mach number in P2 is approximately 2.75 times that in N2. The two flows have similar outlet collision-scale rarefaction but markedly different pressure-gradient, compressibility and acceleration histories.

\begin{landscape}
\begin{figure}[p]
\centering
\includegraphics[width=0.848\linewidth]{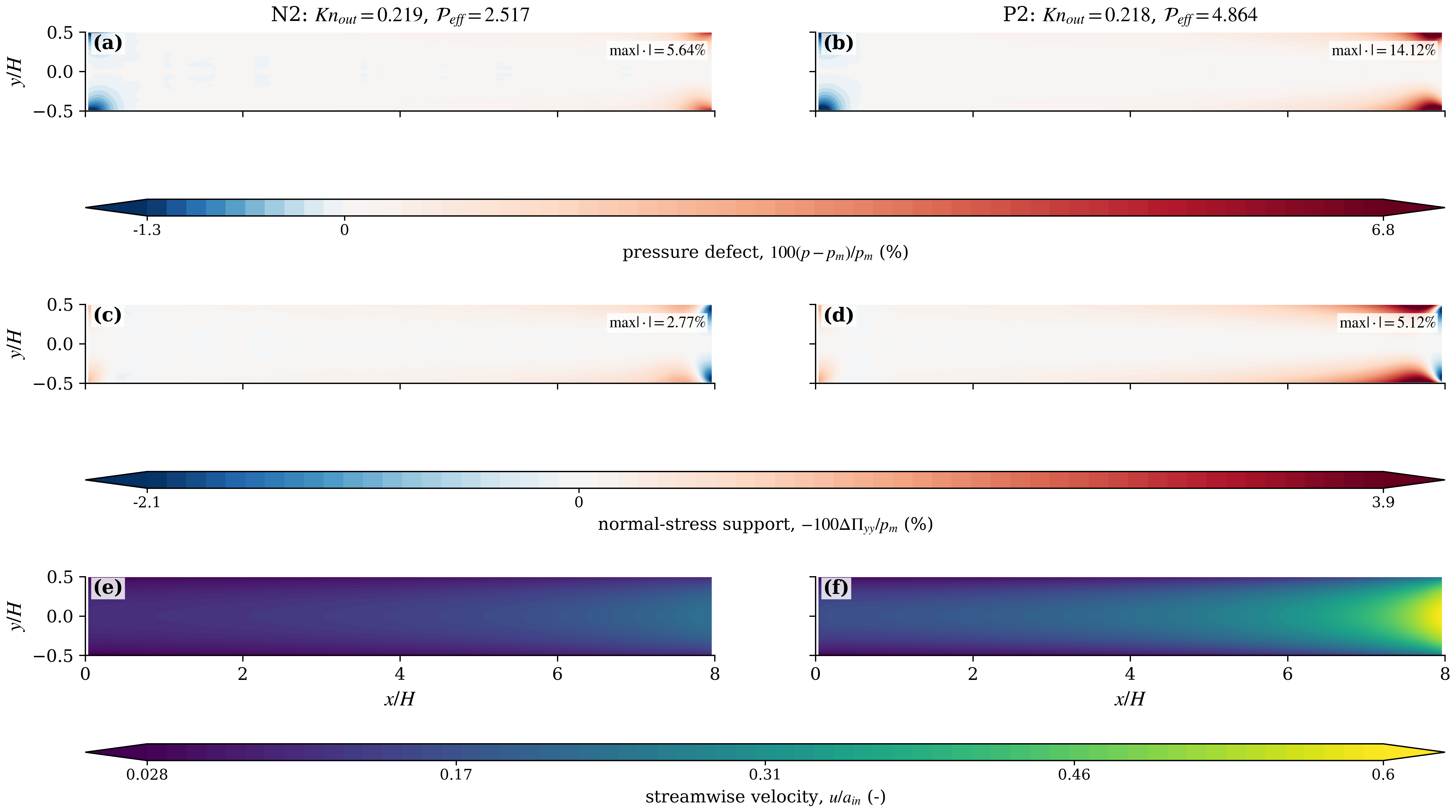}
\caption{Comparison of N2 (left column) and P2 (right column), which have nearly identical outlet rarefaction, $\Kn_{\mathrm{out}}=0.219$ and $0.218$, but substantially different dynamical states. Their realised pressure ratios are $\Peff=2.517$ and $4.864$, and their outlet centreline Mach numbers are $\Ma_{\mathrm{out},c}=0.236$ and $0.649$, respectively. Rows show (a,b) the transverse pressure defect $100(p-p_m)/p_m$, (c,d) the dominant wall-normal normal-stress contribution $-100\Delta\Pi_{yy}/p_m$, and (e,f) $u/a_{in}$, the streamwise velocity normalised by the inlet speed of sound. A common colour scale is used across the two cases within each row. P2 develops a much stronger downstream pressure defect and normal-stress response together with its larger pressure ratio, Mach number and acceleration. The comparison is not a matched-Mach experiment; it demonstrates specifically that matching $\Kn_{\mathrm{out}}$ alone does not determine the transverse-pressure field.}
\label{fig:matched}
\end{figure}
\end{landscape}

The purpose of this comparison is not to isolate pressure forcing while holding the complete dynamical state fixed. Rather, it is a direct counterexample to a one-parameter description based only on $\Kn_{\mathrm{out}}$. The upper row of \cref{fig:matched} shows that the cross-stream pressure defect remains weak over most of N2 but grows strongly towards the downstream wall region in P2. The difference develops progressively along the channel rather than appearing as a spatially uniform multiplication, consistent with the larger pressure ratio and stronger compressible acceleration in P2.

The middle row provides the mechanical interpretation. The field $-\Delta\Pi_{yy}/p_m$ has the same even wall-normal symmetry as the pressure defect and is substantially larger in P2. The greater transverse pressure variation is therefore accompanied by a stronger wall-normal normal-stress response. The difference between the pressure defect and the normal-stress contribution is completed by the streamwise transport of shear stress identified in \cref{fig:ybudget,tab:budget}.

The lower row shows $u/a_{in}$, the streamwise velocity normalised by the inlet speed of sound. It is a velocity proxy rather than the local Mach number, because the local sound speed changes with temperature. Nevertheless, together with the reported $\Ma_{\mathrm{out},c}$ values it makes the dynamical difference explicit: P2 accelerates much more strongly and reaches a considerably larger compressible-flow state than N2.

Quantitatively, the interior-mean wall pressure defect in P2 is \MatchedMeanRatio{} times that of N2, and the ratio rises to \MatchedDownRatio{} at $x/L=0.875$. The conclusion is deliberately limited: matching $\Kn_{\mathrm{out}}$ does not match $\Ma_{\mathrm{out},c}$, the complete Mach field $\Ma(x,y)$, the temperature field, the pressure-gradient distribution or the acceleration history. Outlet Knudsen number alone is therefore not a sufficient state coordinate for the transverse-pressure amplitude.

\begin{figure}[!tbp]
\centering
\includegraphics[width=0.96\textwidth]{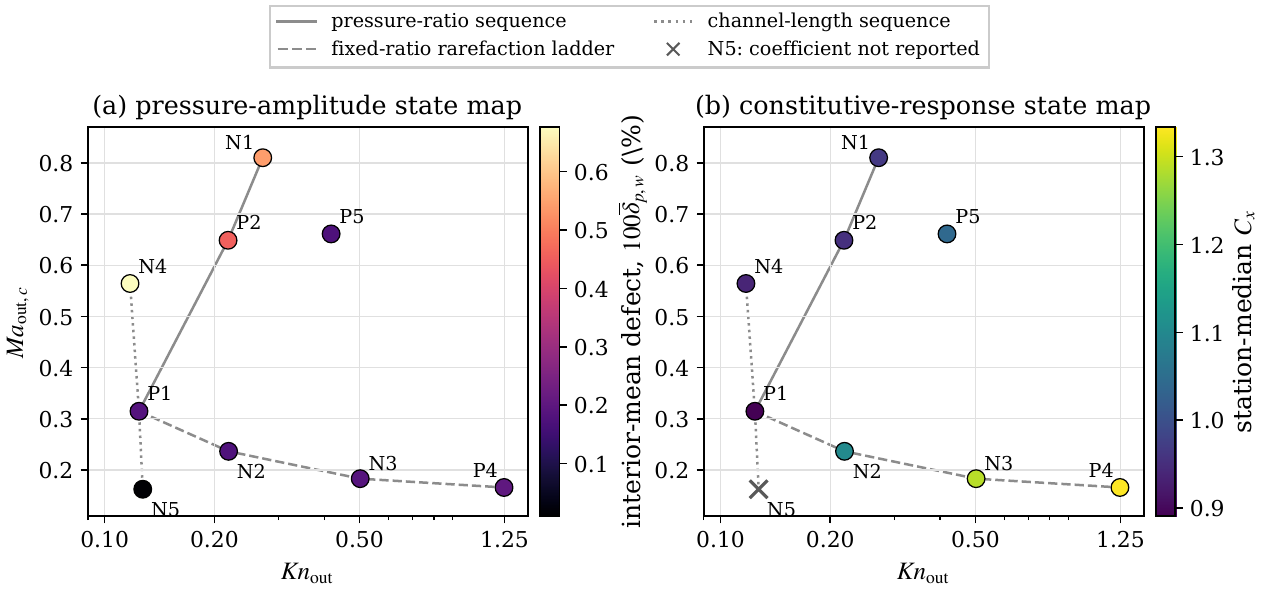}
\caption{Case-level dynamical-state map in the plane of outlet Knudsen number and outlet centreline Mach number. (a) Marker colour denotes the interior-mean near-wall pressure defect. (b) Marker colour denotes the station-median stress coefficient $C_x$; N5 is marked separately because its pressure signal is too weak for a reported coefficient. Grey lines identify the pressure-ratio sequence P1--P2--N1, the fixed-ratio rarefaction ladder P1--N2--N3--P4 and the channel-length sequence N4--P1--N5. The pressure-ratio branch exhibits strong amplitude growth with increasing Mach number, whereas the rarefaction branch exhibits strong constitutive change at nearly constant pressure amplitude. The isolated P5 point also shows that comparable outlet Mach number does not imply comparable pressure amplitude.}
\label{fig:dynamical-map}
\end{figure}

The distinction between a collision-scale coordinate and a dynamical-state coordinate becomes clearer when all cases are placed in the $(\Kn_{\mathrm{out}},\Ma_{\mathrm{out},c})$ plane. Figure~\ref{fig:dynamical-map} colours the same case locations first by the interior-mean pressure defect and then by the station-median stress coefficient. Along the pressure-ratio sequence P1--P2--N1, the branch moves upward to larger Mach number and the pressure amplitude grows strongly, while $C_x$ changes comparatively little. Along the fixed-ratio rarefaction ladder P1--N2--N3--P4, the branch moves towards larger $\Kn$ and smaller Mach number; the pressure amplitude remains nearly constant while $C_x$ increases substantially. The channel-length sequence N4--P1--N5 provides a second near-constant-$\Kn$ contrast with a wide range of Mach number and pressure amplitude.

Mach number is nevertheless not a sufficient amplitude coordinate by itself. P2 and P5 have very similar outlet centreline Mach numbers, 0.649 and 0.661, but their interior-mean defects are 0.453\% and 0.174\%, respectively. Taken together with the streamwise-momentum budget and the forcing
diagnostics, the state map supports a two-part interpretation.  The
pressure amplitude is not organised by either
$\Kn_{\mathrm{out}}$ or $\Ma_{\mathrm{out},c}$ alone, but responds to
their coupled dynamical context, including the local pressure forcing,
compressible acceleration and the channel-length-dependent forcing
history.  By contrast, along the fixed-ratio $\epsilon=8$ ladder,
$\Kn_{\mathrm{out}}$ increases while $\Ma_{\mathrm{out},c}$ decreases
and the pressure amplitude changes only weakly, whereas the
station-median coefficient $C_x$ increases substantially.  This
branchwise contrast associates the constitutive response more directly
with rarefaction, without implying a universal relation
$C_x=C_x(\Kn)$.

\subsection{Local forcing and compressible acceleration organise the pressure amplitude}

\begin{figure}[!tbp]
\centering
\includegraphics[width=0.88\textwidth]{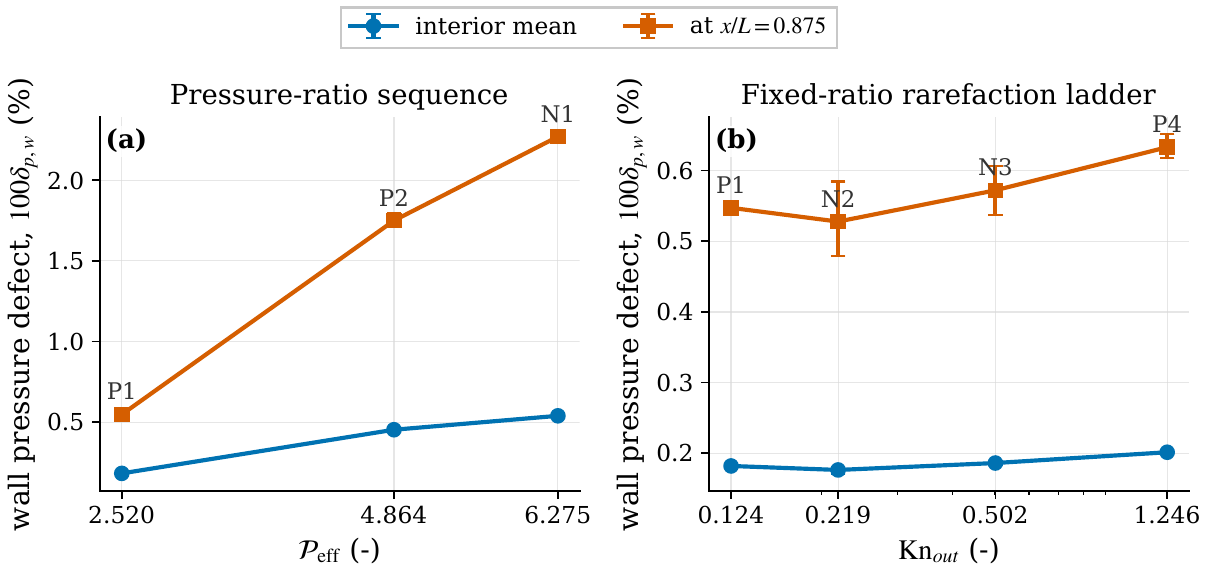}
\caption{Transverse-pressure amplitude along two controlled parameter sequences.  Blue circles denote $\overline{\delta}_{p,w}$, the near-wall pressure defect averaged over the canonical interior $0.1\leq x/L\leq0.9$, where the overbar denotes the streamwise average, and orange squares denote the local value $\delta_{p,w}$ at $x/L=0.875$.  (a) Pressure-ratio sequence P1--P2--N1 at $\eps=8$ and nearly fixed inlet pressure.  Increasing $\Peff$ amplifies both measures and disproportionately strengthens the downstream response.  (b) Fixed-nominal-pressure-ratio ladder P1--N2--N3--P4.  The defect changes only weakly from $\Kn_{\mathrm{out}}=0.124$ to 0.502, indicating that outlet rarefaction is not the primary amplitude coordinate.  P4 is shown as a qualified high-$\Kn$ extension because its amplitude is sensitive to the distance from the outlet pressure boundary.  Error bars are 95\% finite-cluster percentile intervals.}
\label{fig:amplitude}
\end{figure}

The two sequences in \cref{fig:amplitude} contrast collision-scale rarefaction with the coupled effects of pressure forcing, Mach number and streamwise acceleration. The pressure-ratio sequence P1--P2--N1 keeps the channel geometry and inlet pressure nearly fixed while increasing the realised pressure ratio; its Mach-number and acceleration fields necessarily change with that forcing. The fixed-ratio sequence P1--N2--N3--P4 keeps the nominal pressure ratio equal to 2.5 while changing the absolute pressure level and hence $\Kn_{\mathrm{out}}$. The sequences do not isolate Mach number from forcing, because these quantities are dynamically linked in a pressure-driven channel. Their purpose is to determine which part of the resulting state organises the measured amplitude.

In the pressure-ratio sequence, both the interior mean and the downstream defect increase monotonically with $\Peff$.  The increasing separation between the two curves shows that stronger forcing does not amplify the pressure field uniformly: it concentrates the response towards the downstream region, where the pressure gradient, gas acceleration and normal-stress production are largest.

The fixed-ratio ladder behaves differently.  Raising $\Kn_{\mathrm{out}}$ from 0.124 in P1 to 0.502 in N3 changes the interior-mean defect only weakly, and the downstream values remain within a comparatively narrow range.  Thus the resolved rarefaction change does not produce the large amplitude variation caused by increasing pressure ratio.  P4 shows a modest further increase, but its outlet mean free path is comparable with the remaining distance to the pressure boundary; it is therefore treated as a boundary-sensitive high-$\Kn$ extension rather than a fully developed endpoint.

The contrast between the panels is the main physical message of \cref{fig:amplitude}: across the present sequences, the local dynamical state represented by the pressure forcing and compressible acceleration retained in $\Gfull$ better organises the magnitude and streamwise localisation of the transverse pressure field than outlet rarefaction alone. Along the fixed-ratio branch, rarefaction is accompanied by larger changes in the constitutive stress response discussed in \cref{fig:C,fig:clouds,fig:preelim} than in pressure amplitude. Pressure amplitude and stress organisation are therefore related but distinct observables.

\begin{figure}[!tbp]
\centering
\includegraphics[width=0.88\textwidth]{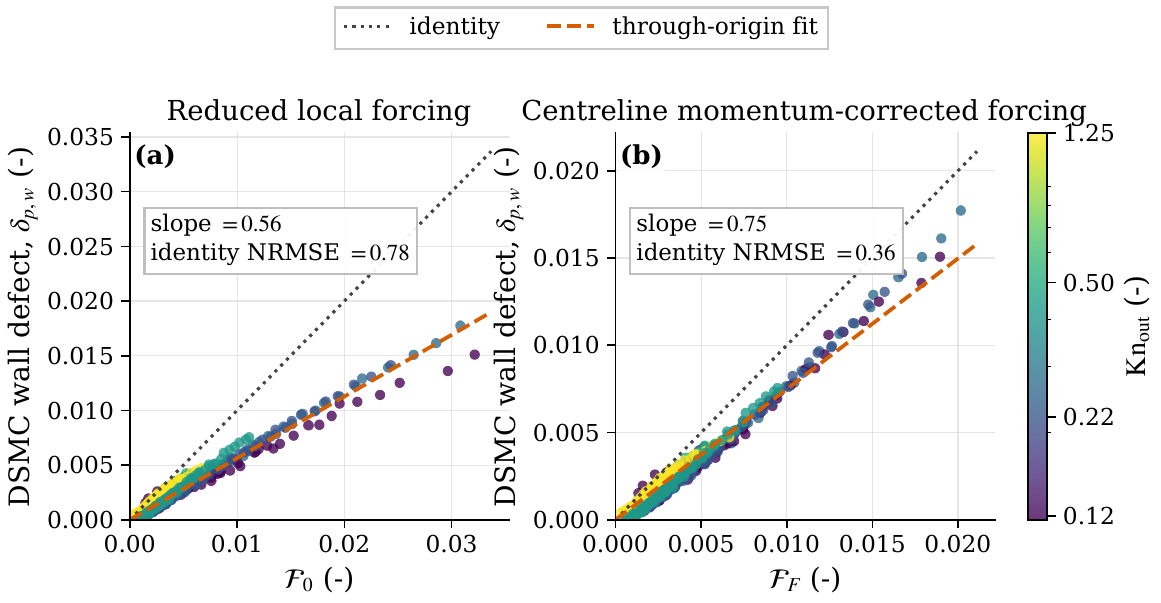}
\caption{Stationwise DSMC wall defect versus the small-defect forcing coordinates $\mathcal F_0=y_w^{*2}\Gzero^2/\Cx$ and $\mathcal F_F=y_w^{*2}\Gfull^2/\Cx$.  Each marker represents a resolved streamwise station and is coloured by case outlet Knudsen number.  The dotted line is the identity relation expected for an exact local reconstruction; the dashed line is a through-origin fit.  (a) The reduced pressure-gradient forcing produces a pronounced multiplicative amplitude bias.  (b) Restoring centreline inertia and $\partial_x\Pi_{xx,m}$ moves the fitted slope towards unity and reduces the identity-based NRMSE, but the remaining sub-unity slope and scatter show that one centreline forcing and one scalar station coefficient do not exhaust the two-dimensional dynamics.  Because all inputs are extracted from the same DSMC fields, the comparison is conditional and in-sample.}
\label{fig:forcing}
\end{figure}

The small-defect form of the tangent law permits a more direct amplitude test. Let $y_w^*=y_w/H$ denote the common dimensionless near-wall sampling location, where $y_w$ is the corresponding dimensional wall-normal coordinate. The reduced and momentum-corrected dimensionless forcing coordinates, $\mathcal F_0$ and $\mathcal F_F$, are defined by
\begin{equation}
\mathcal F_0=\frac{y_w^{*2}\Gzero^2}{\Cx},
\qquad
\mathcal F_F=\frac{y_w^{*2}\Gfull^2}{\Cx}.
\label{eq:forcing-coordinates}
\end{equation}
Here $\delta_{p,w}=\delta_p(x,y_w)$ is the dimensionless near-wall pressure defect and $\mathcal F$ denotes either $\mathcal F_0$ or $\mathcal F_F$. If the local reduced reconstruction were exact, the measured wall defect would satisfy $\delta_{p,w}=\mathcal F$.  Each point in \cref{fig:forcing} represents one selected streamwise station from the combined case set.  The dotted identity line therefore tests absolute amplitude agreement, while the dashed through-origin fit separates multiplicative bias from the remaining scatter.

With the original reduced forcing $\mathcal F_0$, the fitted slope is substantially below unity: the forcing coordinate systematically overestimates the measured pressure amplitude even though the cross-case correlation is high.  Replacing $\Gzero$ by the momentum-corrected $\Gfull$ moves the slope towards unity and approximately halves the identity-based normalised root-mean-square error (NRMSE).  The correction is strongest in P2, N1 and N4, precisely the cases for which supplementary table~S1 identifies a non-negligible centreline inertial contribution.

Supplementary table~S2 reports the through-origin slope, identity NRMSE, optimally rescaled NRMSE and Pearson $R^2$ for both abscissae. Restoring centreline inertia and $\partial_x\Pi_{xx,m}$ moves the slope from 0.562 to 0.744 and reduces the identity-based NRMSE from 0.785 to 0.375. After each coordinate is optimally rescaled, however, the NRMSE changes from 0.149 to 0.160. The correction therefore removes a systematic amplitude bias without establishing a uniquely improved stationwise collapse. Both $\Gfull$ and $\Cx$ are obtained from the DSMC fields being reconstructed, so the comparison is an in-sample mechanistic diagnosis of the reduced forcing.

\subsection{Rarefaction reorganises the stress response}
\begin{figure}[!tbp]
\centering
\includegraphics[width=0.90\textwidth]{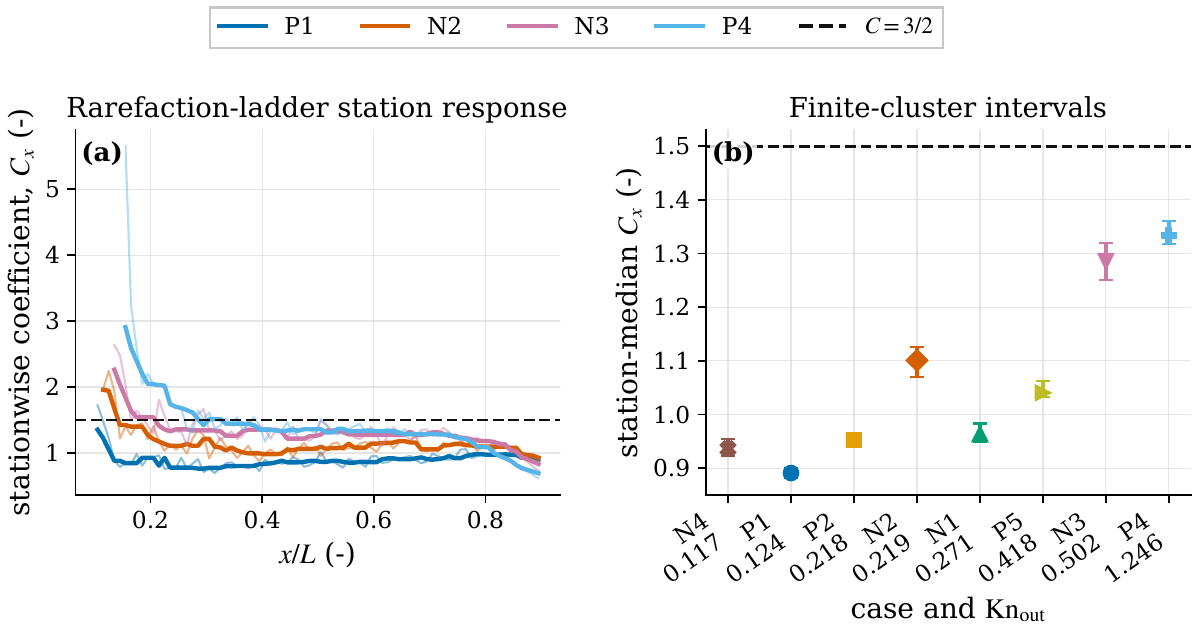}
\caption{Effective stress coefficient along the fixed-ratio rarefaction ladder.  (a) Stationwise $\Cx(x)$ inferred from the selected noise-debiased stress products; thin curves retain the genuine streamwise variability and the occasional excursions produced near the signal threshold.  (b) Case-level station medians with 95\% finite-cluster percentile intervals.  The dashed line marks the reduced NCCR value $C=3/2$.  All resolved cases lie below this value, but the difference between station-balanced and global estimands cautions against interpreting the sequence as a universal monotonic function of $\Kn$.}
\label{fig:C}
\end{figure}

At each streamwise station, $\Cx$ is inferred from the noise-debiased variables in \cref{eq:XY}.  The left panel of \cref{fig:C} displays the stationwise values along the fixed-ratio ladder.  The curves are not expected to be perfectly smooth: the coefficient is a ratio of weak nonlinear stress products, and near the inlet or near the signal-selection threshold small changes in the accepted wall-normal points can produce large local excursions.  These excursions are shown rather than hidden, but the case-level statistic is the median over the canonical station set and is therefore insensitive to isolated near-threshold values.

The right panel shows the corresponding finite-cluster intervals. Every reported coefficient remains below the reduced NCCR value $3/2$. For the independently seeded P cases, the full finite-cluster intervals remain separated from $3/2$ by a margin much larger than the seed-to-seed spread; the conclusion is therefore controlled by the coefficient displacement itself. N5 is not included in the coefficient summary because its interior-mean wall defect is only $0.010\%$ and it is not used in the constitutive conclusions. The station-typical value rises from P1 through N3 and remains high in P4, but the global TLS and OLS slopes reported in \cref{tab:C} do not follow an identical trend. This difference is not a numerical inconsistency: the station median weights every streamwise section equally, whereas a global fit gives greater leverage to larger-amplitude downstream points.

The result does not support one universal value $C=3/2$ over the present pressure-driven family. $\Cx$ varies with streamwise station, case and statistical estimand, so it is used as a diagnostic of the reduced constitutive manifold and as the controlled perturbation in $\mathrm{M}_{C}$, rather than as a proposed material coefficient or a fitted law $C(\Kn)$.

\begin{table}
\centering
\caption{Coefficient estimands for the eight resolved cases used in the constitutive conclusions.  The station median with finite-cluster interval is the primary, station-balanced quantity.  Global total-least-squares (TLS) and ordinary-least-squares (OLS) slopes answer amplitude-weighted questions.  Correlation and interquartile range (IQR) describe the selected point cloud and are reported as observed heterogeneity, not as proven intrinsic constitutive scatter.  N5 is retained in the operating-point table but omitted here because its interior-mean wall defect is only $0.010\%$ and it is not used in the constitutive conclusions.}
\label{tab:C}
\resizebox{\textwidth}{!}{\input{tables/coefficient_summary_generated.tex}}
\end{table}

\begin{figure}[!tbp]
\centering
\includegraphics[width=0.90\textwidth]{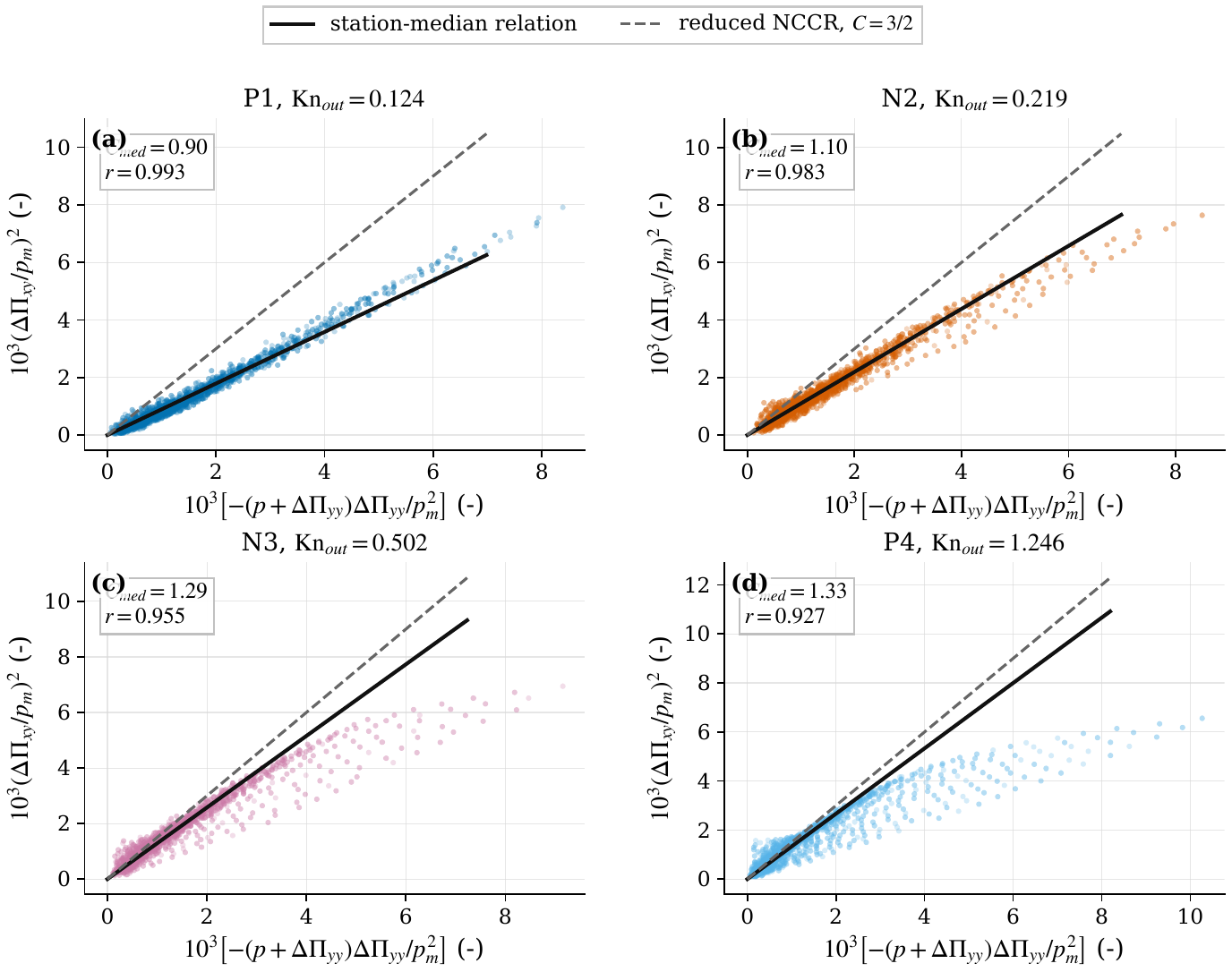}
\caption{Noise-debiased molecular stress relation for P1, N2, N3 and P4.  The abscissa is $10^3[-(p+\Delta\Pi_{yy})\Delta\Pi_{yy}/p_m^2]$ and the ordinate is $10^3(\Delta\Pi_{xy}/p_m)^2$.  Points combine selected wall-normal locations and streamwise stations.  Solid lines use the station-median coefficient and dashed lines use the reduced NCCR value $C=3/2$.  The systematic slope displacement is visible in all cases.  The increasing spread towards P4 is reported as observed heterogeneity; supplementary signal-to-noise matching shows that sampling degradation explains part, but not necessarily all, of the change.}
\label{fig:clouds}
\end{figure}

\Cref{fig:clouds} shows the underlying $X$--$Y$ clouds.  The dashed line is the prescribed relation $Y=(3/2)X$, while the solid line uses the station-median coefficient.  In P1 the selected points form a narrow cloud and the fitted line lies modestly below the NCCR slope.  Through N2 and N3 the preferred slope shifts and the cloud becomes more heterogeneous.  P4 displays the broadest point distribution, although supplementary figure~S1 shows that part of this apparent broadening can be recreated by degrading lower-$\Kn$ cases to the P4 signal-to-noise ratio.

The figure therefore supports two statements of different strength. First, the fixed slope $3/2$ is systematically displaced from the molecular point clouds. Second, the observed relation becomes more heterogeneous with rarefaction. The latter is not labelled intrinsic manifold thickening because square bias, common-centreline covariance, threshold selection and pressure-boundary influence all grow in importance as the signal weakens. Supplementary figure~S1 and the associated signal-to-noise matching procedure distinguish the observed heterogeneity from a stronger claim about an underlying stochastic manifold.

A displaced slope alone would still permit the entire constitutive balance to be repaired by a different common scalar. The next test is stronger: it returns to the pre-elimination tensor equation and asks whether the locally best scalar closes the shear and normal components simultaneously.

\subsection{Componentwise compatibility of the pre-elimination stress balance}
\begin{figure}[!tbp]
\centering
\includegraphics[width=0.92\textwidth]{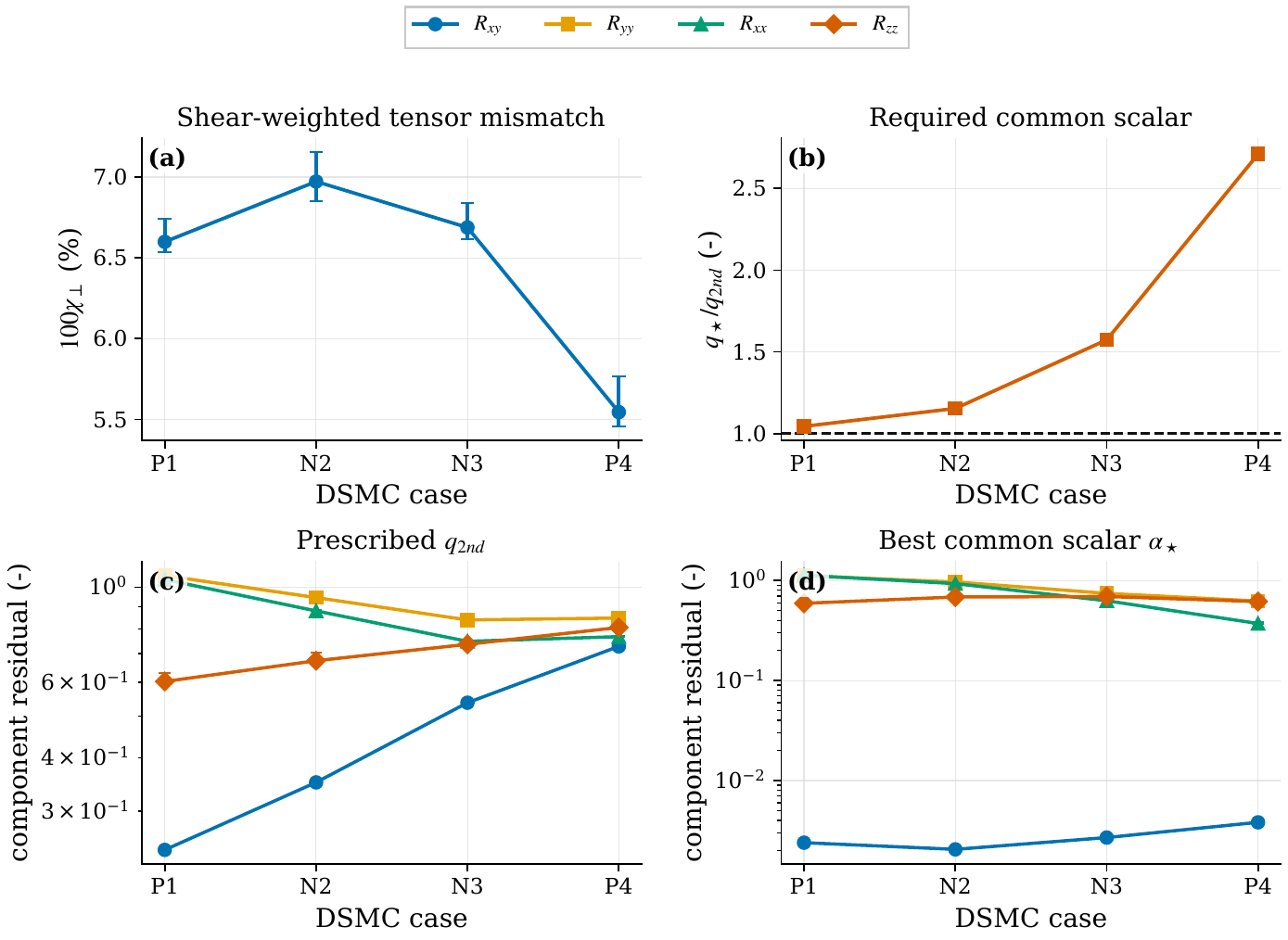}
\caption{Pre-elimination NCCR compatibility along the fixed-ratio rarefaction ladder.  (a) Shear-weighted orthogonal fraction $\chi_\perp$ after fitting the best common scalar.  (b) Scalar relaxation required relative to $q_{2nd}(\kappa)$.  (c) Component residuals $R_{ij}^{(2)}$ obtained with the prescribed $q_{2nd}$, which mix scalar-magnitude and componentwise errors.  (d) Component residuals after applying the best common scalar $\alpha_\star$.  The fitted scalar nearly closes the dominant shear component but leaves substantially larger residuals in the normal components that support the pressure field.  Error bars are 95\% finite-cluster percentile intervals.}
\label{fig:preelim}
\end{figure}

The coefficient inference diagnoses the eliminated manifold, but it does not by itself distinguish a wrong scalar magnitude from a failure of one-scalar component closure. The reduced coefficient is obtained only after eliminating a common scalar from the component equations. A more direct constitutive test is therefore to examine the pre-elimination balance before that cancellation.  \Cref{fig:preelim} separates four questions.  Panel (a) measures the Frobenius-norm tensor mismatch $\chi_\perp$ after fitting the best common scalar.  Its small value shows that the dominant shear-weighted tensor direction remains broadly aligned.  Because shear carries most of the tensor norm, however, this measure alone does not establish closure of the normal components.

Panel (b) shows the scalar ratio $q_\star/q_{2nd}$.  The required scalar increases along the rarefaction ladder even though the modelled second-order factor remains near unity.  Panel (c) reports component residuals with the prescribed $q_{2nd}$; these combine scalar-magnitude error and componentwise incompatibility.  Panel (d) removes the first ambiguity by applying the locally best common scalar $\alpha_\star$.  The shear residual then becomes extremely small, whereas the pressure-carrying normal components retain much larger residuals.  Thus the global tensor alignment is a shear-weighted property and one common scalar does not close the component structure needed by the transverse-pressure reduction. Retaining the complete two-dimensional velocity-gradient tensor does not merely replace $3/2$ by a uniquely modified coefficient: the component elimination then introduces $\Pi_{xx}$, and using the third component to close the relation reintroduces the common scalar. A single kinematically corrected $C$ is therefore not well defined for the full two-dimensional system.

Supplementary table~S7 lists the prescribed-scalar and best-common-scalar residuals side by side. The aggregate normal-subspace residual $R_N^\star$ in \cref{tab:preelim} is 1.105, 0.938, 0.680 and 0.497 for P1, N2, N3 and P4, respectively, while $R_{xy}^\star$ remains 0.002--0.004. Supplementary table~S8 repeats the P4 component test after excluding points within one, two and three local mean free paths of the pressure boundaries. The shear/normal contrast persists under these masks, so it is not created solely by the immediate outlet region, although P4 remains a qualified high-rarefaction endpoint.

\begin{table}
\centering
\caption{Pre-elimination algebraic compatibility. The first column reports $100\chi_\perp$ in percent. Component columns are $R_{ij}^{\star}$ from \cref{eq:Rstar}; $R_N^\star$ is the aggregate normal-subspace residual in \cref{eq:RNstar}. All are evaluated after applying the best common scalar and using the same canonical station selection as the kinematic-decomposition analysis. The small shear residual and much larger normal-subspace residual quantify the componentwise nature of the incompatibility.}
\label{tab:preelim}
\resizebox{\textwidth}{!}{\input{tables/nccr_compatibility_generated.tex}}
\end{table}

The decrease of $R^\star_{yy}$ from the low-$\Kn$ to the high-$\Kn$ end of the ladder should not be read as monotonic closure improvement.  This residual is normalised by $\operatorname{RMS}(K_{yy})$; near the continuum limit the normal component is weak, so a modest absolute mismatch produces a large relative residual.  In contrast, $q_\star/q_{2nd}$ measures the scalar magnitude required by the shear-weighted algebraic balance and is the quantity that grows with rarefaction.  Material transport of the normal stress is too small to explain the low-$\Kn$ residual: over P1, N2, N3 and P4, $\operatorname{RMS}(u\partial_x\Pi_{yy}+v\partial_y\Pi_{yy})/\operatorname{RMS}(K_{yy})$ is $0.0034$, $0.0050$, $0.0064$ and $0.0093$, respectively.

\subsection{Decomposition of the retained kinematic tensor}
\label{sec:kinematic-decomposition}
The compatibility test can be sharpened by separating the retained tensor in \cref{eq:nccr-K} into
\begin{equation}
\bm K=\bm K_{\Pi}+\bm K_p,\qquad
\bm K_{\Pi}=2\devsym{\bm\Pi\boldsymbol\cdot\nabla\bm u},\qquad
\bm K_p=2p\devsym{\nabla\bm u}.
\label{eq:K-decomposition}
\end{equation}
The subscript $\Pi$ labels the nonlinear stress-feedback contribution and the subscript $p$ the pressure--strain contribution. The first term is nonlinear in the existing non-equilibrium stress and represents stress feedback on kinematic production.  The second is the pressure--strain production term that survives in the small-departure limit.  Their relative magnitudes test whether the increasing $q_\star/q_{2nd}$ is associated with growth of nonlinear feedback or with an incompatibility in the dominant pressure--strain/relaxation balance.

Across P1--N2--N3--P4, the median ratio $\norm{\bm K_{\Pi}}/\norm{\bm K}$ is only $0.032$--$0.037$, whereas $\norm{\bm K_p}/\norm{\bm K}$ is $1.000$--$1.004$.  The two terms are weakly opposed: their median cosine changes from approximately $-0.02$ in P1 to $-0.15$ in P4.  Nonlinear stress feedback therefore becomes somewhat more antagonistic with rarefaction but remains far too small to explain the large increase in the required scalar relaxation.

\begin{figure}[!tbp]
\centering
\includegraphics[width=0.92\textwidth]{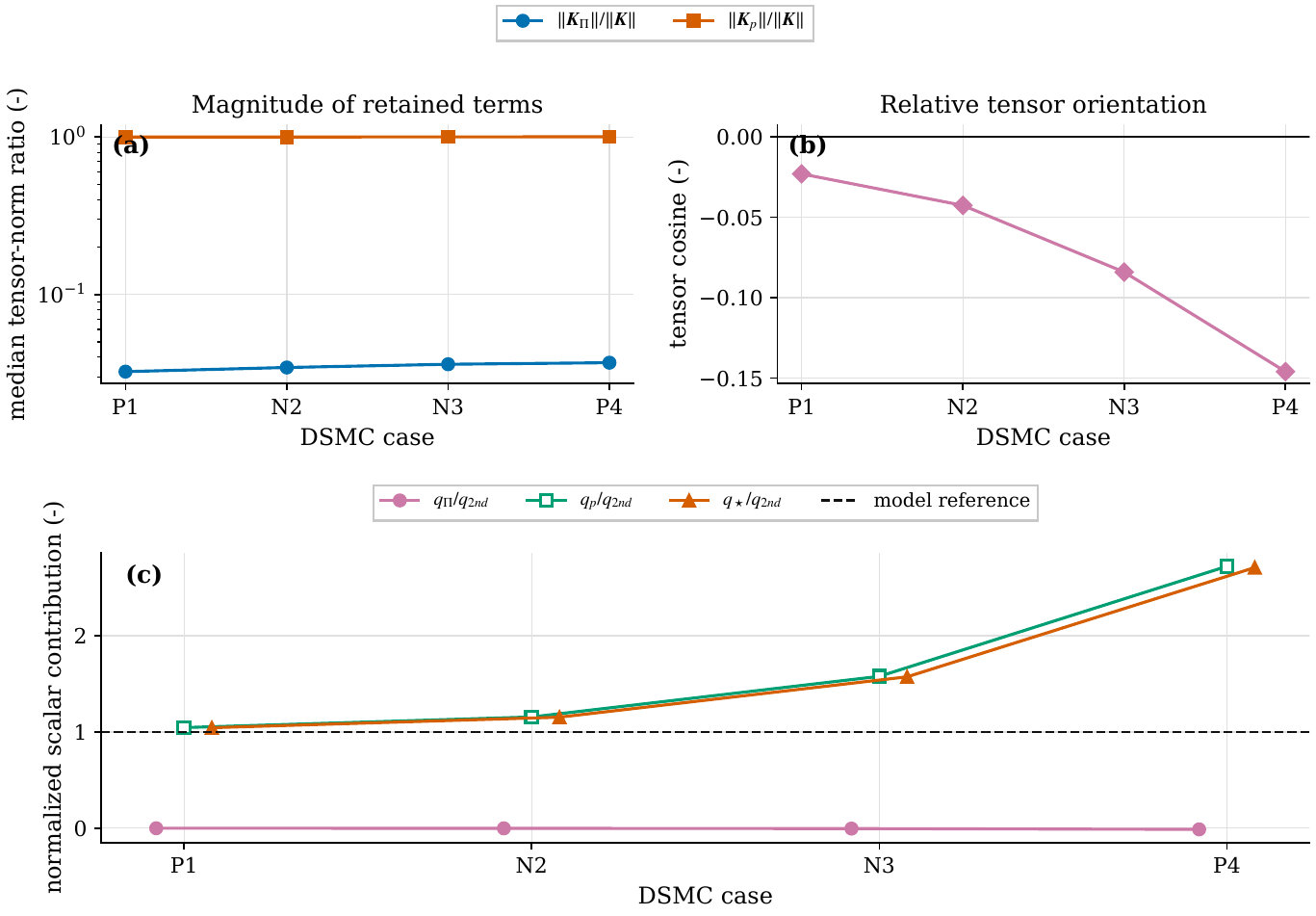}
\caption{Decomposition of the retained pre-elimination NCCR kinematic tensor along the fixed-ratio rarefaction ladder. Panel (a) compares the norms of the nonlinear stress-feedback term $\bm K_{\Pi}$ and pressure--strain term $\bm K_p$ with the total retained tensor $\bm K$. Panel (b) shows their relative tensor orientation. Panel (c) separates the consistently normalized scalar projections $\widehat q_{\Pi}=q_{\Pi}/q_{2nd}$ and $\widehat q_p=q_p/q_{2nd}$ and compares their sum with $q_\star/q_{2nd}$. Pressure--strain production accounts for essentially the entire retained tensor magnitude, whereas nonlinear stress feedback remains only a few percent and weakly opposing. Offset and open markers resolve the close overlap of $\widehat q_p$ and $q_\star/q_{2nd}$.}
\label{fig:kinematic}
\end{figure}

Projecting the contributions separately onto the sampled stress gives
\begin{equation}
\alpha_{\Pi}=\frac{\bm K_{\Pi}:\bm\Pi}{\bm\Pi:\bm\Pi},\qquad
\alpha_p=\frac{\bm K_p:\bm\Pi}{\bm\Pi:\bm\Pi},\qquad
\widehat q_{\Pi}=-\frac{\eta}{p\,q_{2nd}}\alpha_{\Pi},\qquad
\widehat q_p=-\frac{\eta}{p\,q_{2nd}}\alpha_p.
\label{eq:q-decomposition}
\end{equation}
Here $\alpha_{\Pi}$ and $\alpha_p$ are the scalar projections of $\bm K_{\Pi}$ and $\bm K_p$ onto $\bm\Pi$, while $\widehat q_{\Pi}=q_{\Pi}/q_{2nd}$ and $\widehat q_p=q_p/q_{2nd}$ are their normalized relaxation contributions. They obey the pointwise identity $\widehat q_{\Pi}+\widehat q_p=q_\star/q_{2nd}$, so \cref{fig:kinematic} displays a consistently normalized scalar balance. The nonlinear projection $\widehat q_{\Pi}$ is small and negative, while $\widehat q_p$ rises from approximately 1.05 to 2.72 and nearly coincides with $q_\star/q_{2nd}$. Supplementary table~S3 gives the corresponding norm ratios, tensor cosine and normalized scalar projections; separately summarised finite-cluster medians need not add exactly. These results rule out growth of the retained nonlinear stress-feedback term as the source of the scalar deficit. They localise the incompatibility within the closed pressure--strain/relaxation balance, without identifying a unique omitted molecular term.

A common scalar cancels when the component equations are eliminated to form the fixed-$C$ relation.  Consequently, the increase in $q_\star/q_{2nd}$ cannot by itself explain $C\ne3/2$.  The physically decisive evidence is the componentwise residual pattern in \cref{fig:preelim,tab:preelim}: the shear-weighted direction is retained, but the normal components require a different representation.  In the exact molecular stress-moment equation, the inferred discrepancy may include third-order-moment transport and differences between the model collision term and the true collision integral.

\subsection{Compensating constitutive and momentum approximations}
\begin{figure}[!tbp]
\centering
\includegraphics[width=0.88\textwidth]{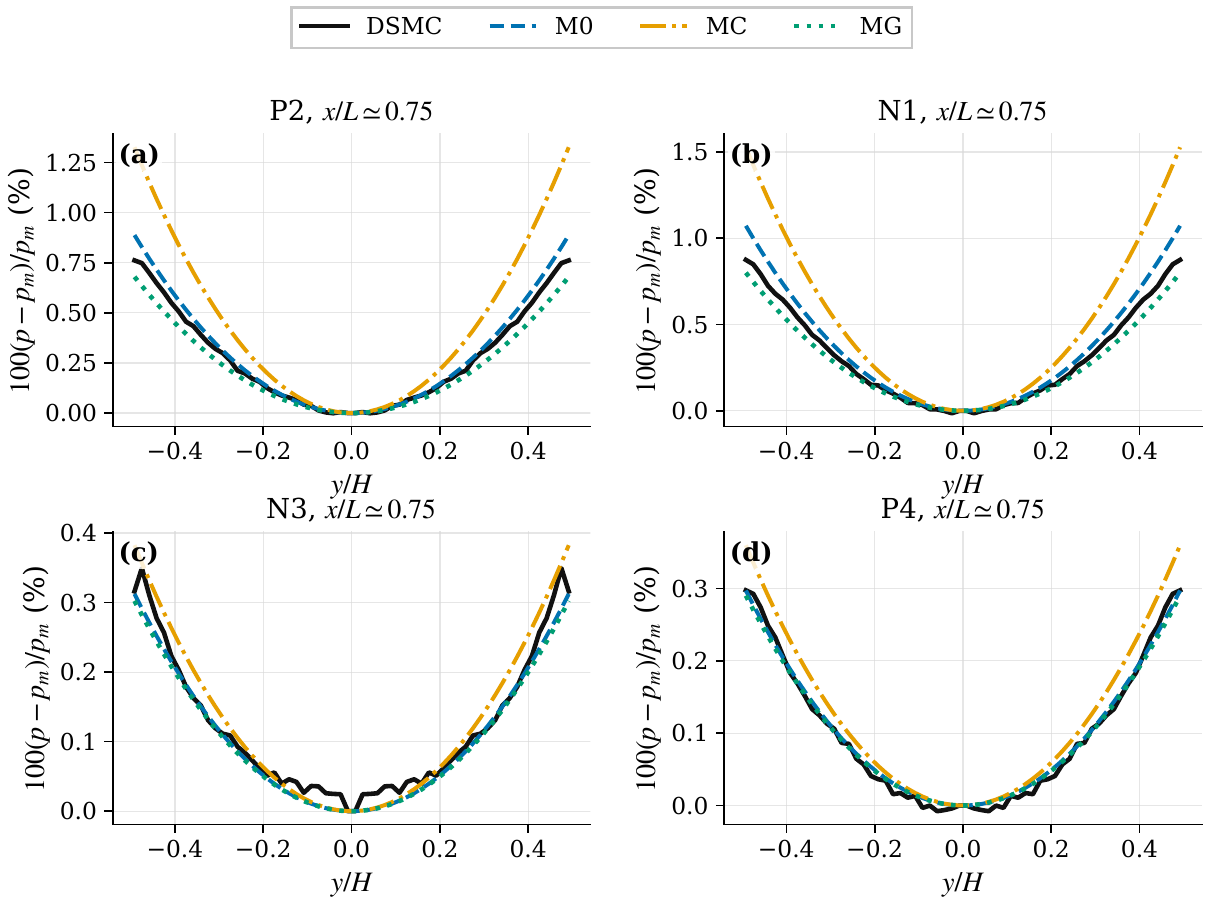}
\caption{Representative cross-stream pressure profiles at $x/L\simeq0.75$.  DSMC is compared with the original reduced reconstruction $\mathrm{M}_{0}$, the coefficient-only substitution $\mathrm{M}_{C}$ and the momentum-forcing substitution $\mathrm{M}_{G}$.  Panels use different ordinate scales because the physical signal varies strongly among cases.  $\mathrm{M}_{C}$ generally increases the defect when $\Cx<3/2$, whereas $\mathrm{M}_{G}$ reduces the strong-forcing overprediction.  $\mathrm{M}_{CG}$ is omitted to avoid over-crowding and is evaluated for every case in \cref{fig:compensation}.}
\label{fig:profiles}
\end{figure}

Representative profiles in \cref{fig:profiles} show how the two reduced ingredients act on the final pressure field.  $\mathrm{M}_{0}$ retains the original coefficient and forcing and captures the convex topology in all panels.  $\mathrm{M}_{C}$ replaces only the coefficient.  Because the inferred $\Cx$ is generally below $3/2$, this substitution increases the tangent argument and therefore amplifies the predicted wall defect; in strong-forcing cases it worsens an existing overprediction.  $\mathrm{M}_{G}$ retains $C=3/2$ but restores centreline inertia and $\partial_x\Pi_{xx,m}$, reducing the effective forcing and often bringing the amplitude closer to DSMC.

\begin{figure}[!tbp]
\centering
\includegraphics[width=0.88\textwidth]{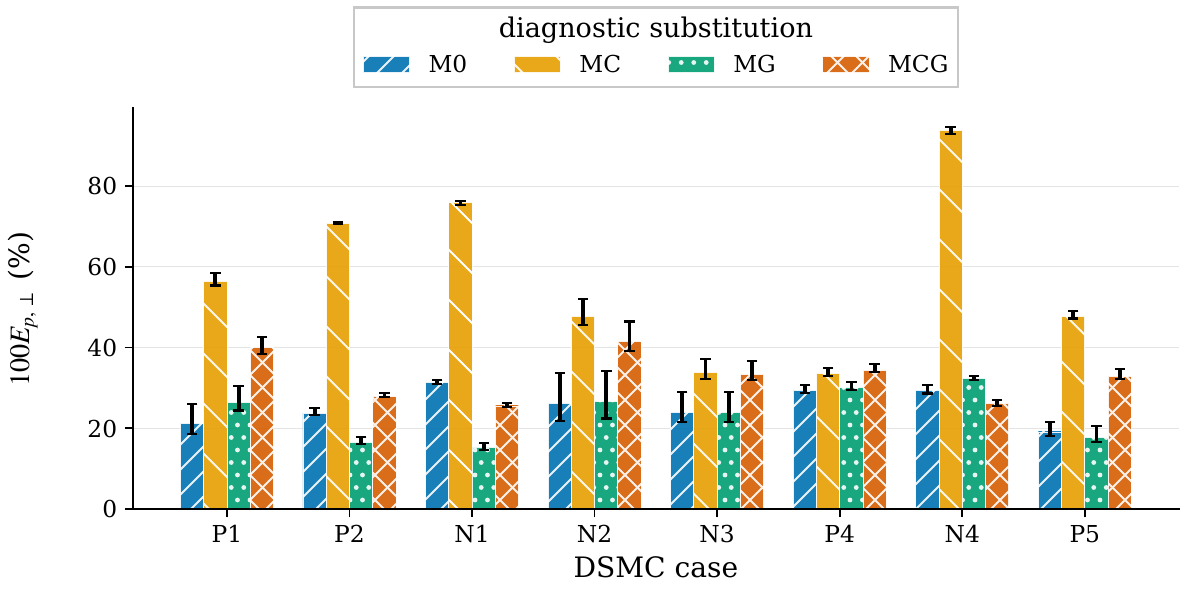}
\caption{Signal-normalised transverse-pressure error for the four canonical DSMC-informed reconstructions.  $\mathrm{M}_{0}$ uses the original coefficient and reduced forcing; $\mathrm{M}_{C}$ replaces only the coefficient; $\mathrm{M}_{G}$ replaces only the streamwise forcing; and $\mathrm{M}_{CG}$ replaces both.  Error bars are 95\% finite-cluster percentile intervals.  The coefficient-only change often worsens the reconstruction, while the forcing correction improves several strong-forcing cases and the combined variant is not universally optimal.  Numerical values are listed in supplementary table~S4.}
\label{fig:compensation}
\end{figure}

The four panels use case-specific ordinate ranges because the transverse signal varies substantially between operating points.  The purpose is to compare model shape and relative amplitude within each case rather than to compare absolute defect magnitudes across panels.  $\mathrm{M}_{CG}$ is omitted from the profile figure to preserve readability; its combined effect is included in the complete error matrix of \cref{fig:compensation}.

The complete case matrix in \cref{fig:compensation} turns these profile observations into a quantitative diagnostic.  The a priori NSF normal stress evaluated on DSMC kinematics is not a solved NSF/slip boundary-value model, but it shows that a local first-order stress does not recover the pressure-support channel.  $\mathrm{M}_{0}$ greatly reduces the flat-pressure error.  In P2 and N1, replacing the forcing alone changes $\Ep$ from \PtwoMzero{}\% to \PtwoMG{}\% and from \NoneMzero{}\% to \NoneMG{}\%, respectively.  By contrast, the coefficient-only substitution often increases the error.  The combined variant is not uniformly best; N4 is the clearest case in which applying both changes is beneficial.

Supplementary table~S4 reports the medians and 95\% finite-cluster percentile intervals for every bar.  It also prevents the graphical comparison from being mistaken for an exact orthogonal error decomposition.  The four variants share DSMC inputs and are conditional substitutions.  Their value is to reveal the directions in which each approximation moves the observable.  The repeated occurrence of opposite movements is the evidence for stress--momentum compensation.

A complementary case-held-out calculation assesses transfer of a single multiplicative amplitude factor while retaining the case-resolved DSMC-informed coordinates $\mathcal F_0$ and $\mathcal F_F$. Across the eight folds, the median error decreases from 19.5\% to 12.8\% and the mean from 18.1\% to 16.2\%. The response is branch dependent: the momentum-informed coordinate improves N3, P4 and N4, whereas P1, P2, N1, N2 and P5 do not improve under the pooled-station fit. Equal total weighting of the training cases preserves the median reduction, from 20.1\% to 12.5\%, with four of eight folds improving. The check therefore identifies useful transfer in selected conditions rather than a universal amplitude coordinate; full foldwise and sequence-extension results are given in supplementary table~S11.

At the reduction level, the two substitutions move the observable in opposite directions. $\mathrm{M}_{C}$ isolates the effect of the constitutive substitution, which increases the pressure amplitude, while $\mathrm{M}_{G}$ isolates the effect of the momentum-forcing substitution, which decreases it in strongly accelerated cases. Their partial cancellation explains why $\mathrm{M}_{0}$ can outperform either isolated substitution. This reduction-level compensation is distinct from, but mechanically consistent with, the exact normal-stress/shear-transport balance identified earlier. The diagnostic matrix is a sensitivity analysis rather than an exact orthogonal error partition.

\subsection{Robustness and range of the molecular diagnostics}
\begin{figure}[!tbp]
\centering
\includegraphics[width=0.92\textwidth]{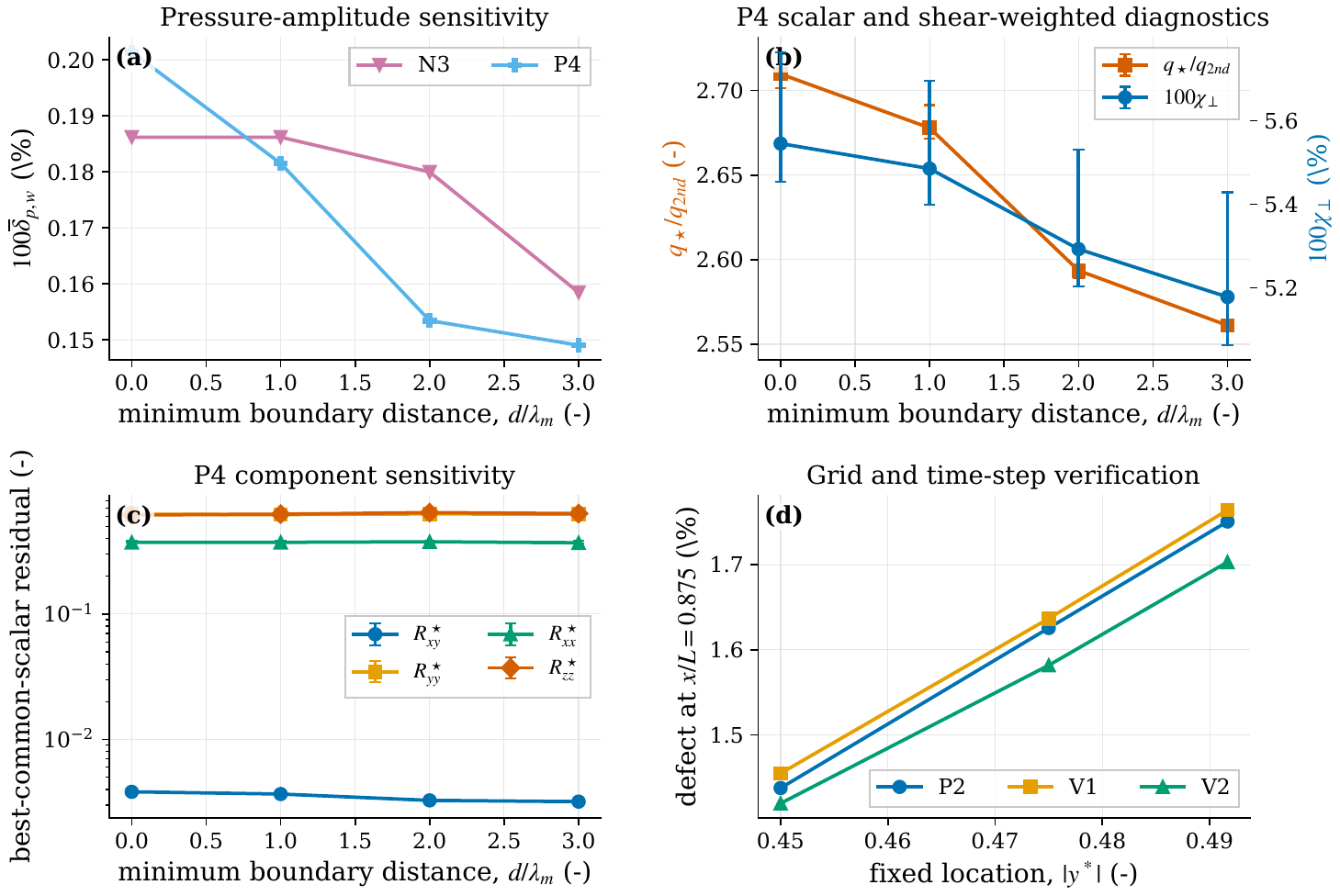}
\caption{Boundary-distance and numerical sensitivity. Here $d=\min(d_{in},d_{out})$ is the distance to the nearer pressure boundary and $\overline{\delta}_{p,w}$ is the streamwise-interior mean near-wall pressure defect. (a) Near-wall pressure amplitude for N3 and P4 as the interior is restricted to points lying more than one, two and three local mean free paths from both pressure boundaries.  (b) P4 station coefficient, scalar-relaxation ratio and shear-weighted tensor mismatch under the same masks.  (c) P4 component residuals after applying the best common scalar; the normal residual remains much larger than the shear residual.  (d) P2 grid and time-step verification at common physical wall-normal locations.  Numerical values and finite-cluster intervals are reported in supplementary tables~S5, S8 and S9.}
\label{fig:sensitivity}
\end{figure}

\Cref{fig:sensitivity} combines the checks most directly connected to the headline conclusions.  Panel (a) replaces the fixed fractional interior by masks requiring one, two and three local mean free paths from both pressure boundaries.  N3 changes little until the most restrictive mask, whereas P4 loses a noticeable part of its measured pressure amplitude as the outlet region is removed.  This confirms that P4 cannot be used as an unqualified fully developed amplitude endpoint.

Panel (b) applies the same masks to the P4 station coefficient, scalar-relaxation ratio and shear-weighted tensor mismatch.  These diagnostics vary less than the pressure amplitude.  Panel (c) resolves the component question: after applying the best common scalar, the shear residual remains very small while the normal-subspace residual remains much larger under every mask.  The componentwise incompatibility is therefore not generated solely by the immediate pressure-boundary region.  Supplementary tables~S5 and S8 provide the numerical values and finite-cluster intervals for these two mask studies.

Panel (d) compares the baseline P2 calculation with the spatially refined case V1 and the half-time-step case V2 at identical physical wall-normal locations.  Both the interior-mean defect and the downstream defect change only by a few percent, far less than the matched P2--N2 contrast.  Supplementary table~S9 lists the corresponding values.  These checks support the principal conclusions while retaining the explicit qualification that a dedicated high-statistics P4 boundary study would be the most valuable additional simulation.

\subsection{Context relative to the published tangent-law validation}
Supplementary table~S6 compares two present operating points with near-matched cases from Appendix~B of \citet{EjtehadiMyong2026}. Because the published molecular fields are unavailable for common-window reprocessing, the comparison is deliberately near matched. The present full-domain $\mathrm{M}_{0}$ errors remain of the same small order as the published full-pressure errors, confirming that the tangent law remains an excellent representation of the total pressure field.

The same table also reports the full-domain signal-relative error and the fraction $\norm{p-p_m}/\norm p$. The transverse signal carries only a few tenths of one percent of the total pressure norm, yet its relative reconstruction error is of order $10^{-1}$. Thus an error of order $10^{-3}$ in the full pressure can coexist with a material error in the specific non-equilibrium correction. The earlier conclusion that the tangent law represents the total pressure accurately therefore remains valid. What changes here is the level of interrogation: the signal norm and internal budgets show that total-field agreement does not validate the constitutive and momentum reductions separately.

\section{Discussion}
\label{sec:discussion}
\subsection{A mechanically complete pressure-support picture}
The exact wall-normal budget changes the interpretation of the convex pressure field from a one-term correspondence to a signed transport balance. $-\Delta\Pi_{yy}$ is the dominant source of wall-normal support, but it generally supplies more than the measured pressure defect. Streamwise transport of shear stress removes the excess, while transverse inertia is small. Because the normal-stress and shear-transport contributions have the same even wall-normal symmetry, the correction modifies amplitude more strongly than topology. This is why the pressure and normal-stress maps can look nearly identical even when the reduced one-term relation has a 20--30\% error on the signal scale.

This distinction would be obscured if the residual were normalised by $p_m$. A discrepancy of order $10^{-3}p_m$ is small relative to the total pressure but substantial relative to a transverse field that is itself only a few tenths of one percent of $p_m$. The combination of the exact budget and the signal-relative norm is therefore essential: one identifies the omitted mechanics, and the other gives those mechanics the correct scale.

\subsection{Dynamical state and rarefaction organise different observables}
The matched P2--N2 experiment shows that $\Kn_{\mathrm{out}}$ is not a sufficient coordinate for the pressure amplitude, but it should not be interpreted as a pure forcing experiment: the two cases also differ strongly in Mach number, temperature and acceleration history. The state map in \cref{fig:dynamical-map} makes this distinction explicit. Along the pressure-ratio branch, increasing Mach number accompanies a large increase in pressure amplitude while the stress coefficient changes little. Along the fixed-ratio rarefaction branch, $\Kn_{\mathrm{out}}$ increases and $\Ma_{\mathrm{out},c}$ decreases, the pressure amplitude changes weakly and the constitutive coefficient changes strongly.

Mach number is a useful state indicator but not an independent control parameter. It is generated jointly by pressure forcing, temperature and channel length, and comparable outlet Mach numbers do not guarantee comparable transverse pressure fields, as the P2--P5 contrast demonstrates. The momentum-based coordinate $\Gfull$ is more mechanistic because it contains the local centreline acceleration and streamwise normal-stress gradient that enter the governing momentum balance. The corresponding ratio $|\Gfull/\Gzero|$, reported in supplementary table~S12, measures how strongly the quasi-one-dimensional forcing is modified in each case.

The earlier short-channel DSMC study of \citet{RoohiDarbandiMirjalili2009} used Mach-number fields and gradient-length breakdown measures to show that rarefaction and compressibility must be considered together when pressure gradients become strong. The analogous local pressure-gradient Knudsen number for the present problem is
\begin{equation}
\Kn_{\mathrm{GLL},p}=\lambda_m\left|\frac{\partial\ln p_m}{\partial x}\right|
=\Kn_m|\Gzero|,
\label{eq:KnGLLp}
\end{equation}
where $\Kn_m=\lambda_m/H$ is the centreline Knudsen number. This quantity combines the local collision scale with the pressure-gradient length and is a natural candidate for future pressure-boundary diagnostics. It is not evaluated in the present analysis and is not used as evidence for the mechanism reported here.

The channel-length sequence should be interpreted within the same framework. Because $H$ is fixed, varying $L/H$ by changing $L$ changes the axial forcing and acceleration histories rather than independently varying wall-normal confinement. A true independent confinement experiment would vary $H$ while matching local forcing, Mach number and rarefaction.

\subsection{What the coefficient audit establishes}
The coefficient inference makes the constitutive layer measurable. The resolved molecular fields do not support one fixed value $C=3/2$ over the present family, while the detailed case trend depends on whether streamwise stations are weighted equally or by stress amplitude. $\Cx$ is therefore interpreted as a local diagnostic of the reduced manifold rather than a material property.

The pre-elimination test gives the stronger result. A best common scalar makes the dominant shear component highly compatible with the retained kinematic tensor, which explains the small shear-weighted Frobenius mismatch. The same scalar leaves much larger residuals in the normal components that carry the cross-stream pressure. The result is therefore not merely an inaccurate value of $3/2$: the one-scalar structure that is adequate for the dominant shear direction does not close the pressure-relevant component subspace.

The kinematic decomposition further shows that this incompatibility is not caused by the retained nonlinear stress-feedback term becoming dominant. Pressure--strain production accounts for essentially all of the retained tensor magnitude, while the nonlinear projection remains small and weakly opposing. The analysis localises where the closed algebraic reduction becomes incompatible, but it does not identify a unique omitted molecular term; third-order-moment transport and collision-model differences remain possible contributors in the exact stress-moment hierarchy.

\subsection{Why the tangent law can remain accurate}
The $\mathrm{M}_{0}$--$\mathrm{M}_{CG}$ matrix connects the mechanical and constitutive findings. A coefficient below $3/2$ increases the tangent argument and tends to amplify the predicted defect. Restoring centreline inertia and $\partial_x\Pi_{xx,m}$ decreases the effective forcing in strongly accelerated cases. The two corrections therefore act in opposite directions, and their partial offset can make the original reduction closer to DSMC than the coefficient-only substitution.

This is the main modelling implication of the study. Agreement of a reduced law with a weak observable is not sufficient evidence that the internal closure mechanics are independently accurate. The exploratory case-held-out check gives lower median and mean errors for the momentum-informed coordinate, although the improvement is concentrated in three of eight pooled-station folds. It therefore provides limited evidence of transfer within parts of the present case family; the equal-case and sequence-extension checks likewise show branch-dependent performance. A predictive extension should determine both $\Gfull$ and $\Cx$ from a closed system and be assessed on independent operating families using the same signal-relative and budget-based diagnostics. Conceptually, this distinction parallels the a priori/a posteriori separation in turbulence modelling, where accurate field predictions can coexist with compensating discrepancies in individual model terms and transfer requires the governing closure structure to remain faithful \citep{DuraisamyIaccarinoXiao2019}.

\subsection{Scope and claim boundaries}
The conclusions concern the fixed-coefficient pressure-driven tangent reduction. The four reconstructions are DSMC-informed diagnostics, and the pre-elimination audit evaluates the closed algebraic NCCR equation on molecular fields. P4 is treated as a boundary-aware high-rarefaction extension, with headline amplitude statements anchored in $\Kn_{\mathrm{out}}\lesssim0.5$; N5 is retained only as a low-amplitude channel-length endpoint and is excluded from constitutive inference.

\section{Conclusions}
\label{sec:conclusions}
The central outcome of this study is not a replacement value for the NCCR coefficient $3/2$. It is a mechanistic explanation of why the reduced tangent law can represent the transverse pressure field even when the two reductions used to derive it are not independently accurate.

The complete wall-normal momentum budget shows that the non-equilibrium normal stress is the dominant support of the convex pressure defect, but it systematically over-supports the measured amplitude. Streamwise transport of shear stress acts with the opposite sign and supplies the leading correction, while transverse inertia is small. This signed balance preserves the pressure topology while modifying its magnitude, explaining why a visually accurate tangent profile can retain a substantial signal-level error.

Controlled case sequences separate the scale of the observable from the organisation of the stress state. Flows with nearly identical outlet Knudsen number develop markedly different pressure amplitudes when their Mach-number and acceleration histories differ. Conversely, P2 and P5 have comparable outlet Mach number but different pressure amplitudes, so Mach number alone is also insufficient. Across the present sequences, pressure amplitude is better organised by the local momentum forcing, including compressible acceleration, while the fixed-ratio rarefaction branch shows larger changes in the molecular stress response and component compatibility than in pressure amplitude.

The constitutive audit shows that the fixed reduced coefficient is not recovered and, more importantly, that the locally best common scalar closes the dominant shear component far more accurately than the normal components responsible for the pressure field. The distinction is therefore componentwise; it cannot be removed by substituting one fitted scalar into an otherwise unchanged law. The pressure--strain term dominates the retained pre-elimination kinematics, so the growing discrepancy is not explained by nonlinear stress feedback becoming large.

The diagnostic reconstruction matrix then exposes the compensation mechanism. Correcting the constitutive coefficient alone often increases the pressure error, while restoring omitted streamwise-momentum terms reduces the amplitude bias in strongly accelerated cases. Their opposing effects allow the original reduced law to remain accurate in the total-pressure norm. The branch-dependent held-out results indicate that the momentum correction is useful in selected conditions but is not yet a universally transferable coordinate. More generally, successful prediction of a weak non-equilibrium observable does not guarantee that the internal closure balances are correct. Predictive improvement requires a closed model for both the stress response and the streamwise forcing, evaluated with complete momentum budgets, held-out conditions and an error norm tied to the observable itself.

\begin{bmhead}[Supplementary material.]
Supplementary material for this article is supplied as a separate file and is included in the submission package.
\end{bmhead}

\begin{bmhead}[Acknowledgements.]
The authors acknowledge the computational resources used to generate and analyse the DSMC data.
\end{bmhead}

\begin{bmhead}[Funding.]
This work was supported by the National Research Foundation of Korea (NRF) Grant (RS2024-00397400) funded by the Ministry of Science and ICT, Republic of Korea.
\end{bmhead}

\begin{bmhead}[Author contributions.]
O.~Ejtehadi and E.~Roohi conceived the study. O.~Ejtehadi performed the DSMC simulations and primary post-processing. E.~Roohi supervised the project, contributed to the analysis strategy and manuscript development, and serves as corresponding author. Both authors interpreted the results and approved the final manuscript.
\end{bmhead}

\begin{bmhead}[Competing interests.]
The authors report no conflict of interest.
\end{bmhead}

\begin{bmhead}[Use of artificial-intelligence tools.]
Generative artificial-intelligence tools were used for language editing and code-development assistance under the authors' supervision. The authors independently checked the analysis, numerical values, figures and scientific interpretation and take full responsibility for the manuscript.
\end{bmhead}

\begin{bmhead}[Data availability statement.]
The article source package contains the manuscript and supplement sources, compiled PDFs, processed summary tables, figure files and scripts for the revision-specific secondary analyses. A separate full reproducibility archive prepared for public deposition contains the executed namelists, raw sampled DSMC outputs, processed fields, complete finite-cluster results and master analysis pipeline. The DSMC solver source and executable are not distributed. A persistent repository identifier will be inserted after the public deposit is finalised.
\end{bmhead}

\bibliographystyle{jfm}
\bibliography{references}

\end{document}


\maketitle

\section{Additional streamwise-momentum and forcing diagnostics}
The complete streamwise-momentum budget and stationwise forcing metrics used in the main paper are listed in \cref{tab:S1,tab:S2}. The displayed ranges are finite-cluster percentile intervals obtained by repeating the differentiation and projection operations within each cluster resample.

\begin{table}
\centering
\caption{RMS terms in the complete streamwise-momentum equation, expressed as percentages of RMS $|\partial_xp|$. Brackets are 95\% finite-cluster percentile intervals. Where both interval limits round to the same displayed value, the interval width is smaller than the printed precision.}
\label{tab:S1}
\resizebox{\textwidth}{!}{\input{supplement_tables/table_S1.tex}}
\end{table}

\begin{table}
\centering
\caption{Stationwise forcing diagnostics. The corrected forcing reduces systematic amplitude bias but is evaluated in sample from the DSMC centreline fields.}
\label{tab:S2}
\input{supplement_tables/table_S2.tex}
\end{table}

\section{Retained-kinematic decomposition and reconstruction matrix}
The main paper presents the kinematic decomposition that localises the incompatibility within the retained pressure--strain/relaxation balance. The corresponding numerical summary is listed in \cref{tab:S3}; the complete $\mathrm{M}_{0}$--$\mathrm{M}_{CG}$ reconstruction matrix is given in \cref{tab:S4}.

\begin{table}
\centering
\caption{Numerical kinematic-tensor decomposition over the canonical selected interior. The scalar projections are normalized by the same $q_{2nd}$ as the reported total $q_\star/q_{2nd}$.}
\label{tab:S3}
\resizebox{\textwidth}{!}{\input{tables/nccr_kinematic_decomposition_generated.tex}}
\end{table}

\begin{table}
\centering
\caption{Signal-normalised pressure errors in percent. M0--MCG entries show the median followed by the 95\% finite-cluster percentile interval. The a priori NSF column is evaluated on DSMC kinematics and has no cluster interval here.}
\label{tab:S4}
\resizebox{\textwidth}{!}{\input{supplement_tables/table_S4.tex}}
\end{table}

\section{Boundary-distance sensitivity}
\Cref{tab:S5} reports the pressure, coefficient and tensor diagnostics after successively excluding points within one, two and three local centreline mean free paths of either pressure boundary.

\begin{table}
\centering
\caption{Mean-free-path-aware sensitivity of pressure and stress diagnostics.}
\label{tab:S5}
\resizebox{\textwidth}{!}{\input{supplement_tables/table_S5.tex}}
\end{table}

\section{Context relative to the published tangent-law errors}
\Cref{tab:S6} compares near-matched operating points with the published Appendix-B full-pressure errors of Ejtehadi \& Myong. The present $E_{p,\perp}^{\mathrm{full}}$ values use the full computational domain; the primary main-text metric is evaluated over the interior domain.

\begin{table}
\centering
\caption{Near-matched comparison with the published full-pressure errors.}
\label{tab:S6}
\resizebox{\textwidth}{!}{\input{supplement_tables/table_S6.tex}}
\end{table}

\section{Best-common-scalar and numerical sensitivity details}
The prescribed second-order scalar residuals and the residuals remaining after applying the best common scalar are listed in \cref{tab:S7}. The P4-only mask sensitivity in \cref{tab:S8} demonstrates that the large separation between shear and normal-subspace compatibility persists as the pressure-boundary exclusion is increased. The fixed-location grid and time-step comparison is listed in \cref{tab:S9}.

\begin{table}
\centering
\caption{Component residuals with the prescribed second-order scalar, $R^{(2)}_{ij}$, and after the best common scalar, $R^*_{ij}$.}
\label{tab:S7}
\resizebox{\textwidth}{!}{\input{tables/nccr_component_residuals_generated.tex}}
\end{table}

\begin{table}
\centering
\caption{P4 tensor diagnostics under increasingly strict boundary-distance masks. Entries are medians with 95\% finite-cluster percentile intervals.}
\label{tab:S8}
\resizebox{\textwidth}{!}{\input{tables/p4_tensor_mask_sensitivity_generated.tex}}
\end{table}

\begin{table}
\centering
\caption{P2 numerical verification at common physical wall-normal locations.}
\label{tab:S9}
\input{tables/verification_generated.tex}
\end{table}

\section{Signal-to-noise matching}
\begin{figure}[tbp]
\centering
\includegraphics[width=0.88\textwidth]{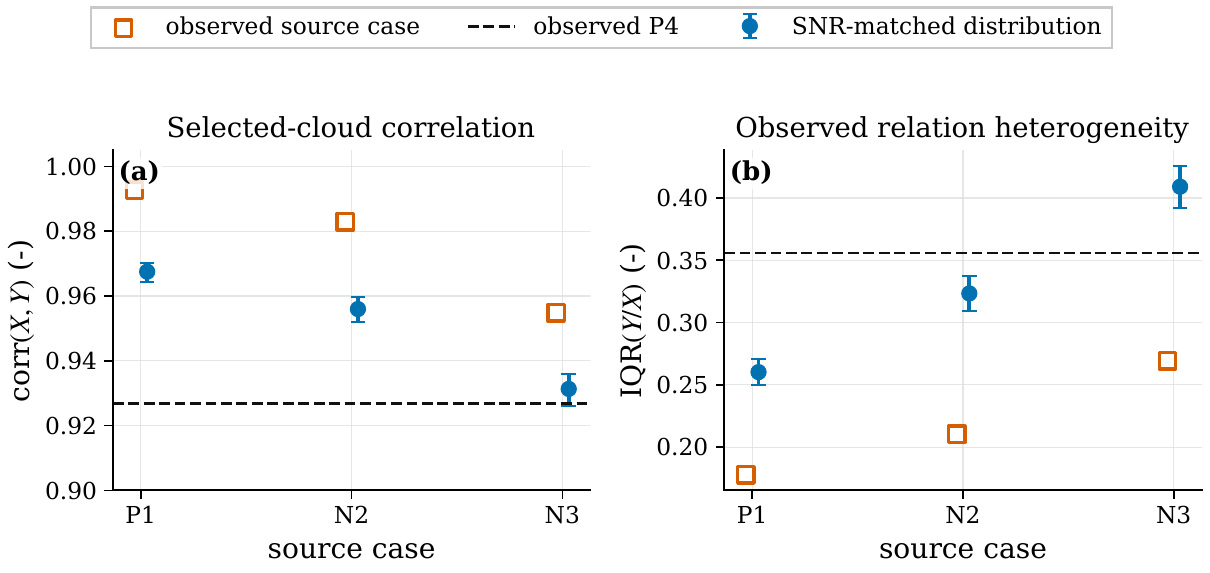}
\caption{Signal-to-noise matching. Open squares are the observed canonical global-cloud estimands, circles and bars are distributions after degradation to the P4 relative-noise covariance, and dashed lines are the observed P4 references.}
\label{fig:snrS}
\end{figure}

The observed correlation and IQR markers use the same global selected cloud and estimands as the coefficient table in the main paper. To test how much heterogeneity can be generated by declining signal-to-noise ratio, the relative-noise vector is defined from the P4 cluster fields as
\begin{equation}
\boldsymbol\epsilon=
\begin{bmatrix}
(X_r-\overline X)/( |\overline X|+X_f)\\[2pt]
(Y_r-\overline Y)/( |\overline Y|+Y_f)
\end{bmatrix},
\qquad
\boldsymbol\epsilon\sim\mathcal N(\boldsymbol 0,\boldsymbol\Sigma_{P4}),
\end{equation}
where $X_f$ and $Y_f$ are the 25th-percentile absolute-signal floors and $\boldsymbol\Sigma_{P4}$ is the pooled covariance of P4 cluster deviations over the canonical interior mask. Each lower-$\Kn$ field is degraded according to
\begin{equation}
X^{(s)}=X(1+\epsilon_X),\qquad Y^{(s)}=Y(1+\epsilon_Y),
\end{equation}
and the canonical selection $0.1\leq x/L\leq0.9$, $|y/H|\geq0.1$, $X^{(s)}>3\,\mathrm{SE}_{JK}(X)$ and $Y^{(s)}\geq0$ is reapplied. Figure~\ref{fig:snrS} shows percentile ranges from 1000 correlated-noise realisations. This is a sensitivity experiment, not a proof that the remaining spread is intrinsic.

\begin{landscape}
\section{Spatial diagnostic maps}
\begin{figure}[H]
\centering
\includegraphics[width=0.97\linewidth]{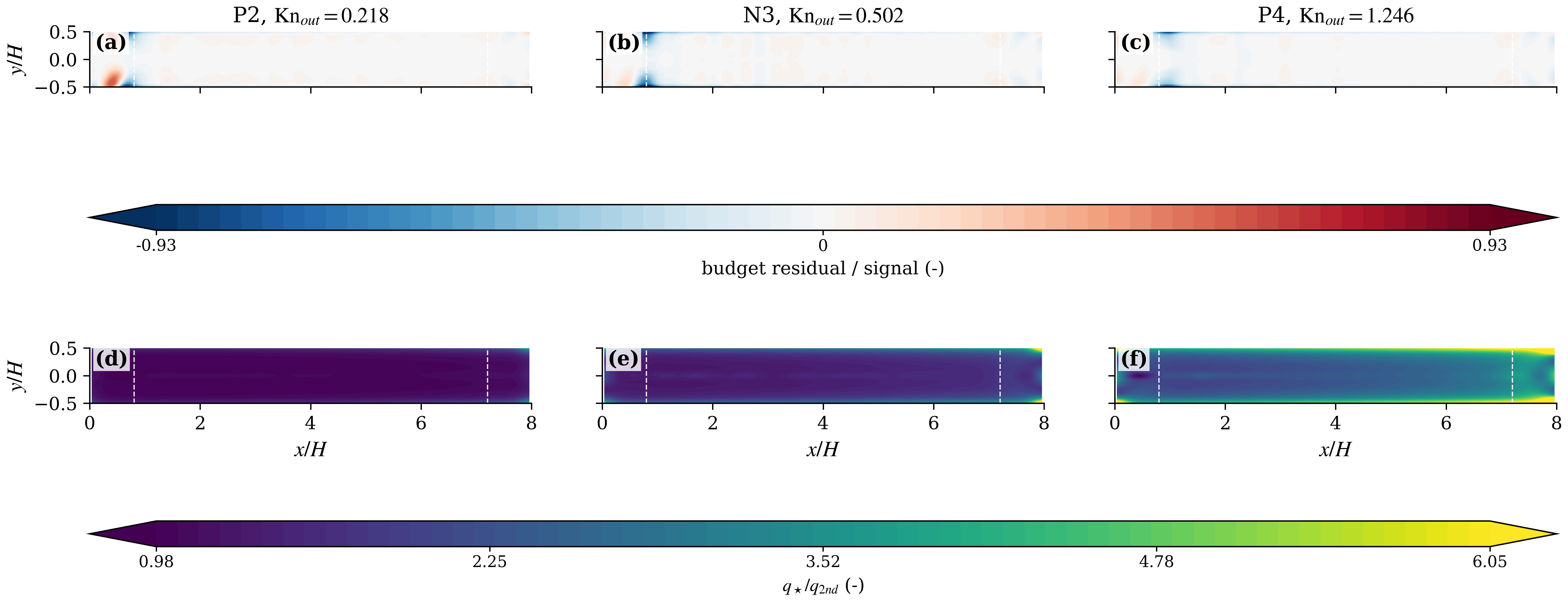}
\caption{Actual-aspect-ratio maps for P2, N3 and P4. Top row: integrated wall-normal momentum residual normalised by the section pressure signal. Bottom row: required-to-modelled scalar relaxation $q_\star/q_{2nd}$. Dashed vertical lines mark the default $0.1\leq x/L\leq0.9$ interior.}
\label{fig:maps1S}
\end{figure}
\end{landscape}

\begin{landscape}
\begin{figure}[H]
\centering
\includegraphics[width=0.97\linewidth]{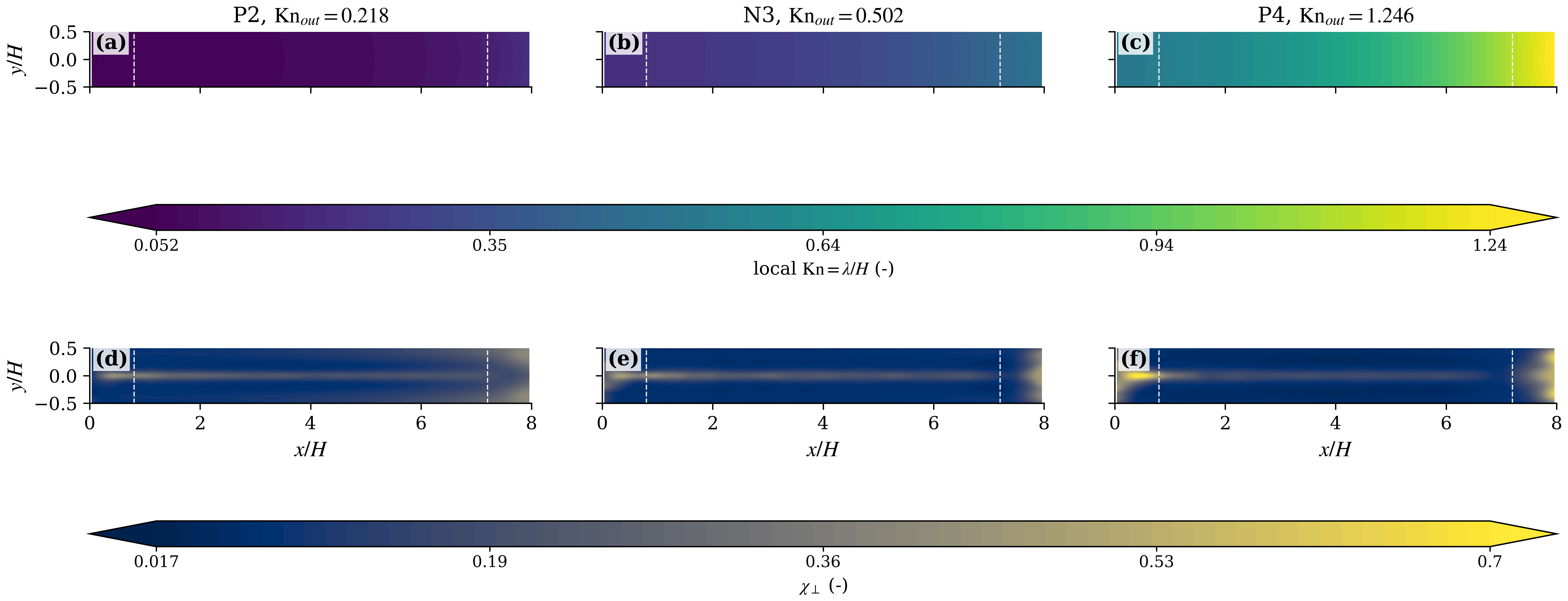}
\caption{Actual-aspect-ratio maps for P2, N3 and P4. Top row: local $\Kn=\lambda/H$. Bottom row: shear-weighted tensor mismatch $\chi_\perp$. Dashed vertical lines mark the default interior.}
\label{fig:maps2S}
\end{figure}
\end{landscape}

\section{Per-case computational settings and statistical units}
\Cref{tab:S10} lists the settings used to form every statistical cluster. The P cases consist of three independent cold-start seeds. The N cases consist of five consecutive, non-overlapping averaging blocks generated by restarting the same calibrated trajectory with fresh sampling accumulators; each block is treated as one finite-cluster unit but is not equated with an independent cold start. ``Averaging steps'' denotes sampled DSMC time steps accumulated after the steady-state transient, not the number of molecular particles contributing to individual cell moments.

\begin{table}
\centering
\caption{Per-case computational and sampling settings. The sampling grid is followed by the number of collision subcells within each sampling cell.}
\label{tab:S10}
\resizebox{\textwidth}{!}{\input{tables/computational_settings_generated.tex}}
\end{table}

The material-transport check discussed in the main text was evaluated over the same canonical interior. The ratios $\operatorname{RMS}(u\partial_x\Pi_{yy}+v\partial_y\Pi_{yy})/\operatorname{RMS}(K_{yy})$ are $0.0034$, $0.0050$, $0.0064$ and $0.0093$ for P1, N2, N3 and P4, respectively; the machine-readable values are supplied in \texttt{normal\_stress\_advection\_check.csv}.

\section{Case-held-out transfer of the amplitude relation}
The stationwise forcing comparison in the main paper is complemented by a case-level leave-one-out test. For a held-out case $h$, a single through-origin amplitude factor is fitted using all stations from the remaining seven resolved cases,
\begin{equation}
\beta_j^{(-h)}=
\frac{\sum_{i\notin h}\mathcal F_{j,i}\,\delta_{p,i}}
{\sum_{i\notin h}\mathcal F_{j,i}^{2}},
\qquad j\in\{0,F\},
\end{equation}
and is then applied without refitting to the omitted case. The transfer error is
\begin{equation}
E_j^{(-h)}=
\frac{\|\beta_j^{(-h)}\mathcal F_j-\delta_p\|_{2,h}}
{\|\delta_p\|_{2,h}}.
\end{equation}
N5 is excluded because its interior-mean wall defect is only $0.010\%$ and it is not used in the constitutive conclusions. The held-out case retains its case-resolved DSMC-informed $C_x$ and forcing coordinate, so the exercise tests transfer of the multiplicative amplitude relation rather than an autonomous closure. Under the pooled-station fit, the median error decreases from 19.5\% to 12.8\% and the mean from 18.1\% to 16.2\%, but only N3, P4 and N4 improve; the other five folds do not. Because cases contribute different numbers of stations, a sensitivity calculation gives every training case equal total weight. It preserves the median reduction, from 20.1\% to 12.5\%, and improves four of eight folds.

Panel (b) of \cref{tab:S11} holds out the non-baseline members of each physical sequence together while retaining P1 as their common reference. The momentum-informed coordinate improves the rarefaction extension and the combined geometry extension, but not the pressure-ratio extension. These extension-level checks are exploratory because the branches are small and share P1; they reinforce the branch-dependent interpretation rather than establishing independent-family transfer.

\begin{table}
\centering
\caption{Transfer of the stationwise amplitude relation while retaining case-resolved DSMC-informed forcing coordinates. (a) Case-level leave-one-out results using the pooled-station training fit. (b) Sequence-extension holds under pooled-station and equal-case weighting; P1 is retained as the common reference. Errors are relative $L_2$ stationwise defect errors in percent.}
\label{tab:S11}
\noindent\textit{(a) Case-level leave-one-out.}\par\smallskip
\resizebox{\textwidth}{!}{\input{tables/heldout_transfer_generated.tex}}
\par\medskip
\noindent\textit{(b) Sequence-extension sensitivity.}\par\smallskip
\resizebox{0.82\textwidth}{!}{\input{tables/heldout_sequence_generated.tex}}
\end{table}

\section{Mach-number and dynamical-state diagnostics}
The outlet centreline Mach number is used in the main paper as a compact descriptor of the compressible flow state, not as an independent control variable. Table~\ref{tab:S12} collects the case coordinates used in the dynamical-state map together with the median forcing-correction ratio
\begin{equation}
\widetilde{\left|\frac{\mathcal G_F}{\mathcal G_0}\right|}
=\operatorname{median}_{x\in\Omega_x}
\sqrt{\frac{\mathcal F_F}{\mathcal F_0}},
\end{equation}
where $\Omega_x$ is the canonical selected streamwise set. The equality follows because $\mathcal F_F$ and $\mathcal F_0$ use the same $y_w^*$ and $C_x$ at each station. Values below unity indicate that restoring centreline inertia and $\partial_x\Pi_{xx,m}$ reduces the effective tangent-law forcing.

The pressure-ratio sequence moves to larger $\Ma_{\mathrm{out},c}$ and larger defect while $C_x$ changes weakly. The fixed-ratio rarefaction ladder moves to larger $\Kn_{\mathrm{out}}$ and smaller $\Ma_{\mathrm{out},c}$ while the defect remains nearly constant and $C_x$ increases. N4 exhibits the largest forcing correction, consistent with its short channel and strong acceleration. P2 and P5 have comparable outlet Mach numbers but different defects, confirming that Mach number alone cannot replace the local forcing coordinate.

\refstepcounter{table}
\label{tab:S12}
\noindent\textbf{Table \thetable.} Case-level rarefaction, Mach-number, pressure-amplitude and constitutive diagnostics. Defect and $C_x$ values are the same finite-cluster medians used in main-paper table~1. The last column is the median magnitude of the centreline momentum-corrected forcing relative to the reduced forcing over the canonical selected stations. A dash denotes a weak-signal case for which the corresponding constitutive or forcing statistic is not used.

\begin{center}
\resizebox{\textwidth}{!}{\input{tables/dynamical_state_summary_generated.tex}}
\end{center}

\section{Bootstrap, estimator and reproducibility details}
For $n$ statistical clusters, the ordinary cluster bootstrap draws $n$ complete fields with replacement. Because $n=3$ or 5, all unique multinomial count vectors are enumerated exactly, giving 10 or 126 weighted resamples. Exact enumeration removes Monte Carlo variability in evaluating this finite-cluster resampling distribution; it does not add independent physical realisations. Every resample repeats field averaging, centreline subtraction, order-two U-statistic formation, signal selection, stationwise fitting, differentiation, integration and pressure reconstruction. The primary coefficient is the median of stationwise through-origin total-least-squares slopes. Global total least squares, ordinary least squares, reverse regression and median $Y/X$ are retained as alternative estimands in the machine-readable tables.

The article source package includes the processed summaries and the following scripts:
\begin{verbatim}
python analysis_heldout_transfer.py
python analysis_mach_dynamics.py
python analysis_revision19_diagnostics.py
\end{verbatim}
They regenerate the revision-specific transfer tables, state-map products, normalized kinematic table and figure. A separate full reproducibility archive prepared for public deposition contains the executed namelists, sampled DSMC fields, finite-cluster outputs and master analysis pipeline. It reproduces the analysis but not the particle simulations because the solver executable and source are not distributed.

%% file: tables/headline_numbers_generated.tex
\newcommand{\MatchedMeanRatio}{2.57}
\newcommand{\MatchedDownRatio}{3.32}

\newcommand{\NormalSupportMin}{1.12}
\newcommand{\NormalSupportMax}{1.30}
\newcommand{\ShearOppMin}{0.14}
\newcommand{\ShearOppMax}{0.32}
\newcommand{\YResidualMin}{0.06}
\newcommand{\YResidualMax}{0.20}

\newcommand{\PtwoMzero}{23.8}
\newcommand{\PtwoMG}{16.6}
\newcommand{\NoneMzero}{31.2}
\newcommand{\NoneMG}{15.2}

\newcommand{\MaxDxSubLambda}{0.77}
\newcommand{\MaxDySubLambda}{0.16}
\newcommand{\MaxDtTau}{0.077}

%% file: tables/operating_points_generated.tex
\begin{tabular}{lrrrrrrrr}
\toprule
Case & $\epsilon$ & $P_{in}$ (kPa) & nominal ratio & $\mathcal P_{eff}$ & $Kn_{out}$ & $M_{out,c}$ & mean defect (\%) & $C_x$ [95\% finite-cluster interval]\\
\midrule
P1 & 8 & 103.10 & 2.5 & 2.520 & 0.124 & 0.315 & 0.182 & 0.891 [0.881, 0.901]\\
P2 & 8 & 103.10 & 5.0 & 4.864 & 0.218 & 0.649 & 0.453 & 0.952 [0.947, 0.965]\\
N1 & 8 & 101.15 & 7.0 & 6.275 & 0.271 & 0.810 & 0.539 & 0.964 [0.948, 0.983]\\
N2 & 8 & 58.70 & 2.5 & 2.517 & 0.219 & 0.236 & 0.176 & 1.101 [1.070, 1.126]\\
N3 & 8 & 25.60 & 2.5 & 2.517 & 0.502 & 0.183 & 0.186 & 1.286 [1.250, 1.320]\\
P4 & 8 & 10.31 & 2.5 & 2.516 & 1.246 & 0.166 & 0.201 & 1.334 [1.317, 1.360]\\
N4 & 4 & 103.10 & 2.5 & 2.502 & 0.117 & 0.564 & 0.676 & 0.936 [0.923, 0.954]\\
N5 & 16 & 103.10 & 2.5 & 2.513 & 0.127 & 0.162 & 0.010 & \multicolumn{1}{c}{--}\\
P5 & 16 & 101.15 & 10.0 & 9.432 & 0.418 & 0.661 & 0.174 & 1.041 [1.032, 1.062]\\
\bottomrule
\end{tabular}

%% file: tables/budget_ci_generated.tex
\begin{tabular}{lrrrr}
\toprule
Case & normal support & shear transport & transverse inertia & residual\\
\midrule
P1 & 1.303 [1.281, 1.311] & -0.324 [-0.328, -0.313] & 0.003 [0.002, 0.003] & 0.162 [0.137, 0.205]\\
P2 & 1.282 [1.278, 1.283] & -0.289 [-0.290, -0.286] & -0.006 [-0.006, -0.006] & 0.067 [0.060, 0.084]\\
N1 & 1.265 [1.261, 1.266] & -0.269 [-0.270, -0.268] & -0.009 [-0.010, -0.009] & 0.064 [0.056, 0.079]\\
N2 & 1.248 [1.214, 1.264] & -0.271 [-0.282, -0.255] & 0.005 [0.005, 0.006] & 0.199 [0.164, 0.264]\\
N3 & 1.159 [1.140, 1.170] & -0.179 [-0.184, -0.170] & 0.008 [0.007, 0.008] & 0.179 [0.153, 0.227]\\
P4 & 1.125 [1.116, 1.127] & -0.136 [-0.137, -0.135] & 0.008 [0.008, 0.008] & 0.140 [0.124, 0.179]\\
N4 & 1.243 [1.238, 1.246] & -0.288 [-0.290, -0.284] & 0.021 [0.020, 0.023] & 0.064 [0.055, 0.083]\\
P5 & 1.241 [1.233, 1.244] & -0.244 [-0.245, -0.240] & -0.006 [-0.006, -0.006] & 0.134 [0.121, 0.161]\\
\bottomrule
\end{tabular}

%% file: tables/coefficient_summary_generated.tex
\begin{tabular}{lrrrrrr}
\toprule
Case & station median [95\% finite-cluster interval] & station IQR & global TLS & global OLS & correlation & IQR$(Y/X)$\\
\midrule
P1 & 0.891 [0.881, 0.901] & [0.831, 0.949] & 0.922 & 0.918 & 0.993 & 0.178\\
P2 & 0.952 [0.947, 0.965] & [0.892, 1.011] & 1.057 & 1.055 & 0.999 & 0.170\\
N1 & 0.964 [0.948, 0.983] & [0.903, 1.042] & 1.099 & 1.097 & 0.999 & 0.172\\
N2 & 1.101 [1.070, 1.126] & [1.036, 1.172] & 1.050 & 1.042 & 0.983 & 0.210\\
N3 & 1.286 [1.250, 1.320] & [1.210, 1.387] & 1.101 & 1.074 & 0.955 & 0.269\\
P4 & 1.334 [1.317, 1.360] & [1.276, 1.455] & 1.003 & 0.957 & 0.927 & 0.356\\
N4 & 0.936 [0.923, 0.954] & [0.914, 0.967] & 0.926 & 0.923 & 0.993 & 0.103\\
P5 & 1.041 [1.032, 1.062] & [0.982, 1.105] & 1.153 & 1.151 & 0.998 & 0.209\\
\bottomrule
\end{tabular}

%% file: tables/nccr_compatibility_generated.tex
\begin{tabular}{lccccccc}
\toprule
Case & $100\chi_\perp$ (\%) & $q_\star/q_{2nd}$ & $R^\star_{xy}$ & $R^\star_{yy}$ & $R^\star_{xx}$ & $R^\star_{zz}$ & $R^\star_N$\\
\midrule
P1 & 6.60 [6.54, 6.74] & 1.044 [1.042, 1.048] & 0.002 [0.002, 0.002] & 1.117 [1.115, 1.119] & 1.130 [1.129, 1.131] & 0.592 [0.576, 0.622] & 1.105 [1.104, 1.107]\\
N2 & 6.97 [6.85, 7.15] & 1.154 [1.152, 1.159] & 0.002 [0.002, 0.002] & 0.967 [0.963, 0.973] & 0.931 [0.929, 0.935] & 0.685 [0.667, 0.727] & 0.938 [0.935, 0.943]\\
N3 & 6.69 [6.62, 6.84] & 1.573 [1.568, 1.580] & 0.003 [0.003, 0.003] & 0.744 [0.740, 0.752] & 0.628 [0.625, 0.633] & 0.693 [0.681, 0.716] & 0.680 [0.677, 0.687]\\
P4 & 5.55 [5.45, 5.76] & 2.710 [2.701, 2.723] & 0.004 [0.004, 0.004] & 0.623 [0.616, 0.644] & 0.372 [0.370, 0.381] & 0.616 [0.608, 0.648] & 0.497 [0.490, 0.515]\\
\bottomrule
\end{tabular}

%% file: supplement_tables/table_S1.tex
\begin{tabular}{lcccc}
\toprule
Case & inertia & $\partial_x\Pi_{xx}$ & $\partial_y\Pi_{xy}$ & residual\\
\midrule
P1 & 4.8 [4.8,4.8] & 0.8 [0.8,0.8] & 94.5 [94.5,94.6] & 2.6 [2.3,3.0]\\
P2 & 10.4 [10.4,10.4] & 1.8 [1.8,1.8] & 91.3 [91.3,91.3] & 1.9 [1.8,2.1]\\
N1 & 12.0 [12.0,12.0] & 2.3 [2.3,2.3] & 90.5 [90.5,90.5] & 1.8 [1.7,2.0]\\
N2 & 2.9 [2.9,2.9] & 1.0 [1.0,1.0] & 97.1 [97.1,97.1] & 2.4 [2.0,3.1]\\
N3 & 1.8 [1.8,1.8] & 1.2 [1.2,1.2] & 98.3 [98.3,98.3] & 1.9 [1.6,2.4]\\
P4 & 1.5 [1.5,1.5] & 1.3 [1.3,1.4] & 98.4 [98.4,98.4] & 1.5 [1.4,1.7]\\
N4 & 15.2 [15.2,15.2] & 2.3 [2.3,2.4] & 86.7 [86.7,86.7] & 2.1 [2.0,2.3]\\
P5 & 4.8 [4.8,4.8] & 1.0 [1.0,1.0] & 95.7 [95.7,95.7] & 1.9 [1.8,2.2]\\
\bottomrule
\end{tabular}

%% file: supplement_tables/table_S2.tex
\begin{tabular}{lrrrr}
\toprule
Forcing diagnostic & slope & identity NRMSE & scaled NRMSE & Pearson $R^2$\\
\midrule
reduced local forcing & 0.562 & 0.785 & 0.149 & 0.966\\
centreline momentum-corrected forcing & 0.744 & 0.375 & 0.160 & 0.985\\
\bottomrule
\end{tabular}

%% file: tables/nccr_kinematic_decomposition_generated.tex
\begin{tabular}{lrrrrrr}
\toprule
Case & $\|K_\Pi\|/\|K\|$ & $\|K_p\|/\|K\|$ & $\cos(K_\Pi,K_p)$ & $q_\Pi/q_{2nd}$ & $q_p/q_{2nd}$ & $q_\star/q_{2nd}$\\
\midrule
P1 & 0.0324 & 1.0002 & -0.0231 & -0.0012 & 1.0455 & 1.0443\\
N2 & 0.0344 & 1.0007 & -0.0427 & -0.0021 & 1.1559 & 1.1538\\
N3 & 0.0361 & 1.0020 & -0.0841 & -0.0048 & 1.5782 & 1.5734\\
P4 & 0.0369 & 1.0043 & -0.1460 & -0.0130 & 2.7227 & 2.7098\\
\bottomrule
\end{tabular}

%% file: supplement_tables/table_S4.tex
\begin{tabular}{lrrrrr}
\toprule
Case & a-priori NSF & M0 & MC & MG & MCG\\
\midrule
P1 & 102.4 & 21.2 [18.6, 26.0] & 56.3 [55.4, 58.4] & 26.4 [24.4, 30.6] & 39.9 [38.5, 42.6]\\
P2 & 99.9 & 23.8 [23.4, 25.1] & 70.7 [70.5, 71.1] & 16.6 [16.1, 17.8] & 27.9 [27.7, 28.8]\\
N1 & 102.0 & 31.2 [30.9, 31.9] & 75.8 [75.4, 76.3] & 15.2 [14.7, 16.4] & 25.5 [25.3, 26.2]\\
N2 & 136.8 & 26.0 [21.8, 33.7] & 47.7 [45.6, 52.1] & 26.7 [22.4, 34.1] & 41.4 [39.1, 46.5]\\
N3 & 287.9 & 23.9 [21.5, 28.9] & 33.8 [32.3, 37.1] & 24.0 [21.5, 28.9] & 33.4 [31.9, 36.7]\\
P4 & 709.7 & 29.4 [28.8, 30.7] & 33.5 [33.0, 35.0] & 30.1 [29.5, 31.4] & 34.4 [33.9, 35.9]\\
N4 & 94.2 & 29.4 [28.6, 30.7] & 93.8 [93.0, 94.6] & 32.3 [31.9, 33.0] & 26.1 [25.5, 27.1]\\
P5 & 123.1 & 19.1 [18.0, 21.7] & 47.6 [47.2, 49.0] & 17.6 [16.5, 20.5] & 32.9 [32.2, 34.7]\\
\bottomrule
\end{tabular}

%% file: supplement_tables/table_S5.tex
\begin{tabular}{lrrrrrrr}
\toprule
Case & cutoff $d/\lambda_m$ & stations & mean defect (\%) & median $C_x$ & residual/signal & $q_\star/q_{2nd}$ & $100\chi_\perp$ (\%)\\
\midrule
N3 & 0 & 80 & 0.1862 & 1.289 & 0.151 & 1.571 & 6.63\\
N3 & 1 & 80 & 0.1862 & 1.289 & 0.151 & 1.571 & 6.63\\
N3 & 2 & 79 & 0.1800 & 1.290 & 0.155 & 1.566 & 6.63\\
N3 & 3 & 74 & 0.1585 & 1.314 & 0.170 & 1.552 & 6.63\\
P4 & 0 & 80 & 0.2014 & 1.333 & 0.124 & 2.710 & 5.45\\
P4 & 1 & 77 & 0.1815 & 1.338 & 0.133 & 2.672 & 5.40\\
P4 & 2 & 62 & 0.1535 & 1.362 & 0.150 & 2.589 & 5.20\\
P4 & 3 & 46 & 0.1491 & 1.362 & 0.151 & 2.550 & 5.06\\
\bottomrule
\end{tabular}

%% file: supplement_tables/table_S6.tex
\begin{tabular}{lrrrr}
\toprule
Case & published $E_{full}$ (\%) & present $E_{full}$ (\%) & present $E_{p,\perp}^{full}$ (\%) & signal/full pressure (\%)\\
\midrule
P1 & 0.211 & 0.206 & 18.6 & 0.260\\
P5 & 0.121 & 0.097 & 18.0 & 0.202\\
\bottomrule
\end{tabular}

%% file: tables/nccr_component_residuals_generated.tex
\begin{tabular}{lrrrrrrrr}
\toprule
Case & $R^{(2)}_{xy}$ & $R^{(2)}_{yy}$ & $R^{(2)}_{xx}$ & $R^{(2)}_{zz}$ & $R^*_{xy}$ & $R^*_{yy}$ & $R^*_{xx}$ & $R^*_{zz}$\\
\midrule
P1 & 0.244 & 1.065 & 1.042 & 0.603 & 0.002 & 1.117 & 1.130 & 0.592\\
N2 & 0.350 & 0.946 & 0.880 & 0.674 & 0.002 & 0.967 & 0.931 & 0.685\\
N3 & 0.538 & 0.839 & 0.748 & 0.736 & 0.003 & 0.744 & 0.628 & 0.693\\
P4 & 0.727 & 0.847 & 0.767 & 0.805 & 0.004 & 0.623 & 0.372 & 0.616\\
\bottomrule
\end{tabular}

%% file: tables/p4_tensor_mask_sensitivity_generated.tex
\begin{tabular}{lrrrrr}
\toprule
Case & cutoff $d/\lambda_m$ & $q_\star/q_{2nd}$ & $100\chi_\perp$ (\%) & $R^*_{xy}$ & $R^*_{N}$\\
\midrule
P4 & 0 & 2.710 (2.701--2.723) & 5.55 (5.45--5.76) & 0.004 (0.004--0.004) & 0.497 (0.490--0.515)\\
P4 & 1 & 2.678 (2.672--2.691) & 5.49 (5.40--5.70) & 0.004 (0.004--0.004) & 0.498 (0.491--0.517)\\
P4 & 2 & 2.594 (2.589--2.607) & 5.29 (5.20--5.53) & 0.003 (0.003--0.004) & 0.506 (0.499--0.526)\\
P4 & 3 & 2.561 (2.549--2.578) & 5.18 (5.06--5.43) & 0.003 (0.003--0.003) & 0.500 (0.493--0.522)\\
\bottomrule
\end{tabular}

%% file: tables/verification_generated.tex
\begin{tabular}{lrrr}
\toprule
Dataset & $|y^*|$ & interior mean defect (\%) & defect at $x/L=0.875$ (\%)\\
\midrule
P2 & 0.450000 & 0.41844 & 1.43759\\
P2 & 0.475000 & 0.46204 & 1.62558\\
P2 & 0.491667 & 0.45276 & 1.75071\\
V1 & 0.450000 & 0.41505 & 1.45489\\
V1 & 0.475000 & 0.45890 & 1.63677\\
V1 & 0.491667 & 0.46853 & 1.76415\\
V2 & 0.450000 & 0.41768 & 1.41957\\
V2 & 0.475000 & 0.46715 & 1.58180\\
V2 & 0.491667 & 0.45554 & 1.70310\\
\bottomrule
\end{tabular}

%% file: tables/computational_settings_generated.tex
\begin{tabular}{lccccccc}
\toprule
Case & sampling grid & collision subcells & initial particles/cell & $\Delta t$ (s) & statistical units & averaging steps/unit & total averaging steps\\
\midrule
P1 & 100\(\times\)60 & 2\(\times\)2 & 53 & 1.0e-11 & 3 seeds & 200,000 & 600,000\\
P2 & 100\(\times\)60 & 2\(\times\)2 & 90 & 1.0e-11 & 3 seeds & 200,000 & 600,000\\
N1 & 100\(\times\)60 & 2\(\times\)2 & 110 & 1.0e-11 & 5 blocks & 90,000 & 450,000\\
N2 & 100\(\times\)60 & 2\(\times\)2 & 53 & 1.0e-11 & 5 blocks & 80,000 & 400,000\\
N3 & 100\(\times\)60 & 2\(\times\)2 & 53 & 2.0e-11 & 5 blocks & 80,000 & 400,000\\
P4 & 100\(\times\)60 & 2\(\times\)2 & 53 & 5.0e-11 & 3 seeds & 200,000 & 600,000\\
N4 & 50\(\times\)60 & 2\(\times\)2 & 53 & 1.0e-11 & 5 blocks & 80,000 & 400,000\\
N5 & 200\(\times\)60 & 2\(\times\)2 & 53 & 1.0e-11 & 5 blocks & 140,000 & 700,000\\
P5 & 200\(\times\)60 & 2\(\times\)2 & 165 & 1.0e-11 & 3 seeds & 200,000 & 600,000\\
\bottomrule
\end{tabular}

%% file: tables/heldout_transfer_generated.tex
\begin{tabular}{lrrrr}
\toprule
Held-out case & $\beta_0^{(-h)}$ & $100E_0^{(-h)}$ (\%) & $\beta_F^{(-h)}$ & $100E_F^{(-h)}$ (\%)\\
\midrule
P1 & 0.565 & 14.4 & 0.756 & 28.4\\
P2 & 0.560 & 6.3 & 0.747 & 12.3\\
N1 & 0.559 & 5.4 & 0.735 & 13.2\\
N2 & 0.562 & 19.2 & 0.752 & 20.2\\
N3 & 0.560 & 27.5 & 0.748 & 12.0\\
P4 & 0.559 & 29.8 & 0.747 & 10.1\\
N4 & 0.589 & 19.7 & 0.745 & 7.9\\
P5 & 0.560 & 22.3 & 0.755 & 25.0\\
\midrule
Median & -- & 19.5 & -- & 12.8\\
Mean & -- & 18.1 & -- & 16.2\\
\bottomrule
\end{tabular}

%% file: tables/heldout_sequence_generated.tex
\begin{tabular}{llrr}
\toprule
Held-out extension & weighting & $100E_0$ (\%) & $100E_F$ (\%)\\
\midrule
pressure-ratio & pooled stations & 6.4 & 13.4\\
 & equal case & 9.6 & 12.7\\
rarefaction & pooled stations & 26.9 & 14.4\\
 & equal case & 28.7 & 14.5\\
geometry & pooled stations & 19.7 & 14.8\\
 & equal case & 19.4 & 10.5\\
\bottomrule
\end{tabular}

%% file: tables/dynamical_state_summary_generated.tex
\begin{tabular}{lrrrrrrr}
\toprule
Case & $\epsilon$ & $\mathcal P_{\mathrm{eff}}$ & $\mathit{Kn}_{\mathrm{out}}$ & $\mathit{Ma}_{\mathrm{out},c}$ & mean defect (\%) & $C_x$ & $\widetilde{|\mathcal G_F/\mathcal G_0|}$\\
\midrule
P1 & 8 & 2.520 & 0.124 & 0.315 & 0.182 & 0.891 & 0.956\\
P2 & 8 & 4.864 & 0.218 & 0.649 & 0.453 & 0.952 & 0.926\\
N1 & 8 & 6.275 & 0.271 & 0.810 & 0.539 & 0.964 & 0.924\\
N2 & 8 & 2.517 & 0.219 & 0.236 & 0.176 & 1.101 & 0.975\\
N3 & 8 & 2.517 & 0.502 & 0.183 & 0.186 & 1.286 & 0.989\\
P4 & 8 & 2.516 & 1.246 & 0.166 & 0.201 & 1.334 & 0.994\\
N4 & 4 & 2.502 & 0.117 & 0.564 & 0.676 & 0.936 & 0.859\\
N5 & 16 & 2.513 & 0.127 & 0.162 & 0.010 & -- & --\\
P5 & 16 & 9.432 & 0.418 & 0.661 & 0.174 & 1.041 & 0.976\\
\bottomrule
\end{tabular}